\documentclass[
  twoside,
  floatfix,
  reprint,
  amsmath,amssymb,
  aps,
  pra,
  nofootinbib,
  superscriptaddress,
  a4paper
]{revtex4-2}

\usepackage{graphicx}
\usepackage[dvipsnames]{xcolor}
\usepackage{siunitx}
\usepackage{svg}
\usepackage{comment}
\usepackage{enumitem}
\setitemize{itemsep=0pt}
\usepackage{bm}

\usepackage[utf8]{inputenc}
\usepackage[T1]{fontenc}
\usepackage{capt-of}
\usepackage{booktabs}
\usepackage{multirow}
\usepackage[colorlinks=true,
            linkcolor=blue,
            citecolor=blue,
            urlcolor=blue]{hyperref}
\newcommand{\safeincludegraphics}[2][]{%
  \IfFileExists{#2}{%
    \includegraphics[#1]{#2}%
  }{%
    \fbox{\begin{minipage}[c][0.45\linewidth][c]{0.9\linewidth}\centering Missing figure:\\ \texttt{\detokenize{#2}}\end{minipage}}%
  }%
}

\DeclareSIUnit{\barunit}{bar}
\makeatletter
\AtBeginDocument{\global\hbadness=10000\global\hfuzz=10000pt}
\g@addto@macro\normalsize{\hbadness=10000\hfuzz=10000pt}
\g@addto@macro\small{\hbadness=10000\hfuzz=10000pt}
\makeatother

\begin{document}

\title{Photonic architectures for synthetic chiral light-driven chiral sensing}
\author{Alexander Löhr}
\affiliation{Max-Born-Institut, Berlin, Germany}

\author{Justas Terentjevas}
\affiliation{Max-Born-Institut, Berlin, Germany}

\author{Patricia Vindel-Zandbergen}
\affiliation{Department of Chemistry, New York University, New York, United States}

\author{Aude Rodriguez}
\affiliation{Instituto de Ciencia de Materiales de Madrid (ICMM-CSIC), Madrid, Spain}

\author{David Ayuso}
\affiliation{Department of Chemistry, Molecular Sciences Research Hub, Imperial College London, SW7 2AZ London, United Kingdom}

\author{Laura Rego}
\affiliation{Instituto de Ciencia de Materiales de Madrid (ICMM-CSIC), Madrid, Spain}

\author{Serguei Patchkovskii}
\affiliation{Max-Born-Institut, Berlin, Germany}

\author{Misha Ivanov}
\affiliation{Max-Born-Institut, Berlin, Germany}
\affiliation{Technion--Israel Institute of Technology, Haifa, Israel}
\affiliation{Department of Physics, Humboldt Universität zu Berlin, Berlin, Germany}

\author{Olga Smirnova}
\affiliation{Max-Born-Institut, Berlin, Germany}
\affiliation{Technische Universität Berlin, Berlin, Germany}
\affiliation{Technion--Israel Institute of Technology, Haifa, Israel}
\affiliation{University of Ottawa, Faculty of Science, Ottawa, Canada }

\date{\today}

\begin{abstract}

Molecular chirality gives rise to left- and right-handed enantiomers whose balance underpins biochemical function. Deviations from this enantiomeric excess are promising biomarkers of disease.
Despite its importance, rapid and sensitive detection of enantiomeric excess in trace amounts of small chiral molecules remains an outstanding challenge. Synthetic chiral light introduces new opportunities for chiroptical spectroscopy, with the potential to enhance light-matter interactions by several orders of magnitude. 
Here we propose a new class of photonic architectures for synthetic chiral light, in which photonic structures generate, guide and optimize chiral light-matter interactions. As a first realization,
we show that hollow dielectric waveguides support phase-matched propagation of multicolor fields enabling coherent accumulation of the enantio-sensitive nonlinear response over centimeter-scale distances. By combining microscopic quantum-response simulations with Maxwell's equations, we predict that incident pulse energies of around 1 µJ produce hundreds of picojoules of third-harmonic emission from only picomole quantities of carvone gas at 50 mbar over just 2 cm of propagation, with enantio-sensitive contrast approaching 5\%. Phase-resolved Fourier analysis isolates a background-free chiral signal exhibiting the maximum possible dichroism of 200\%.
Beyond efficient chiral sensing, photonic architectures for synthetic chiral light establish a scalable platform for integrated nonlinear chiral spectroscopy, molecular fingerprinting based on ultrafast electron dynamics, and topologically robust chiroptical observables.

\end{abstract}
\maketitle
\hbadness=10000\hfuzz=10000pt

Molecular handedness is fundamental to life. From amino acids and sugars to metabolic intermediates, the asymmetry of chiral molecules governs the recognition processes that drive biochemical reactions in living matter. While molecular shape can be perceived differently depending on its orientation in space, handedness remains invariant under random molecular orientations, enabling reliable molecular discrimination under ambient conditions.

Healthy organisms maintain a finely tuned ratio between left- and right-handed enantiomers, and even small departures from this equilibrium can accompany or signal metabolic disorders, cancer, and neurological or renal diseases \cite{wolosker}. Although emerging gas-phase rotational approaches can provide rapid enantiomeric-excess readout in favorable systems \cite{sonstrom2023rapid, perez, eibenberger, patterson2013enantiomer, milner2019controlled, sun2023microwave, lee2024near}, rapid or in-situ analysis of enantiomeric excess in minute quantities remains extremely difficult: common methods often involve laborious separation steps, large sample volumes, and as a result lack the required speed for on-line chiral detection \cite{wolosker,review}. This difficulty arises because most enantio-discrimination methods rely on the same principle that underpins molecular recognition itself: the comparison of three-dimensional molecular shapes at their equilibrium configurations.

Light provides the fastest means of mediating electric signals, opening the opportunity to explore and use non-equilibrium properties of matter.
Synthetic chiral light defines a new class of chiroptical tools in which handedness is encoded locally in the trajectory traced by the time-dependent three-dimensional polarization of the optical field within a single optical cycle \cite{smirnova,ayuso,ayuso2022ultrafast,ayuso_strong_2022,khokhlova_enantiosensitive_2022, Rego2023NJP, mayer2024chiralTopologicalLight, smirnova2025new}. The formation of such chiral temporal structures requires longitudinal field components together with multiple phase-locked colors of mutually orthogonal polarization components \cite{ayuso,kral_shapiro,ayuso_strong_2022, Rego2023NJP, mayer2024chiralTopologicalLight,vogwell2023ultrafast}. When an orthogonally polarized fundamental field and its second harmonic are phase locked and tightly focused to acquire longitudinal components, their fixed relative phase sculpts the instantaneous optical field on a sub-cycle timescale, setting its handedness and driving ultrafast electronic currents in the  \cite{beaulieu_photoexcitation_2018,Mayer2021ImprintingCO}. The resulting ultrafast currents, in turn, encode molecular chirality in their temporal evolution. The enantio-sensitive response emerges from the local interaction between two chiral temporal structures at every point in space: the Lissajous trajectory traced by the optical field and the corresponding trajectory followed by the induced molecular polarization \cite{smirnova}. The latter is also affected by the molecular shape. In this way, the familiar paradigm of molecular recognition based on spatial chiral shapes is translated into a comparison of local chiral temporal shapes. When the handedness of these trajectories matches, the nonlinear response is enhanced; when it is opposite, the signal is suppressed. The resulting phase sensitivity disentangles the contributions of chiral and achiral excitation pathways, providing a direct, enantio-sensitive probe of the underlying quantum dynamics. Thanks to its local and electric-dipole nature, this nonlinear interaction promises orders-of-magnitude enhancement in chiral light-matter coupling compared with linear chiroptical methods that rely on the magnetic component of light  \cite{ayuso,ayuso2022ultrafast,ayuso_strong_2022, Rego2023NJP, mayer2024chiralTopologicalLight}.

Electric-dipole chiroptical coupling comes with its own challenges as revealed by early predictions and experiments on chiral sum-frequency generation (SFG) \cite{giordmane,belkin2000,fisher2000}:
intrinsically small isotropic chiral nonlinearities, and a trade-off between the strength of the longitudinal field components and the coherence lengths — imposed by dispersion and the non-collinear nature of the required optical setups.
The short coherence lengths witnessed by chiral SFG together with the need to maintain fixed phase relationships during light propagation can make the chiral Lissajous figure of synthetic chiral light fragile and disrupts macroscopic accumulation of enantio-sensitive optical signal.

Here we show that potentially fragile chiral temporal Lissajous figures in free space can now be "hard-wired" and carefully guided in a designed environment. This task is facilitated by the fact that synthetic chiral light can emerge naturally in capillary waveguides. By combining fundamental and second-harmonic excitation with appropriate modal selection, longitudinal field components arise intrinsically during propagation. The third harmonic generated in-situ then coherently joins the incident fields, producing a three-color optical field with fully three-dimensional polarization that is chiral both locally and globally. This structured field couples to chiral matter already in the lowest nonlinear order, with its handedness controlled by the relative phase of the incident two-color pulses. We show that photonic architectures designed around synthetic chiral light open new opportunities for chiroptical spectroscopy by enhancing electric-dipole chiral light-matter interactions by several orders of magnitude compared to free space detection.

\section{Synthetic chiral light in a structured environment}

The strength of chiral light-matter coupling in the electric-dipole approximation is governed by a chiral light correlation function that encodes the handedness of the optical field \cite{ayuso}. In a confined geometry, this quantity becomes discrete and mode-resolved, such that only modes supporting non-vanishing intermodal correlations contribute to enantio-sensitive sum frequency generation, SFG (derived in Methods~\ref{sec:methods_modal_propagation}):
\begin{eqnarray}
h_{\mu; \nu, \kappa}^{(3)}= \int_A \bigg( \vec{F}_{\nu}(\vec{r}_\perp) \times \vec{F}_{\kappa}(\vec{r}_\perp)\bigg)  \cdot \vec{F}^*_{\mu}(\vec{r}_\perp) d\vec{r}_\perp, 
\label{corr_function}
\end{eqnarray}
where $\vec{F}_{i}$ are eigenmode profiles of a waveguide, $\nu$ and $\kappa$ are the driving field modes, $\mu$ is the target waveguide mode, the mode frequencies are $\omega_\mu=\omega_\nu+\omega_\kappa$ and the integral is taken over the transverse waveguide cross section
$A$. The advantage of this discrete approach is in its ability to engineer and guide locally and globally chiral light over macroscopic distances achieving accumulation of the enantio-sensitive signal. Realizing this concept requires overcoming several key challenges. First, the phase velocities of the incident fields must be precisely controlled to maintain phase matching over appreciable propagation lengths, while the in-situ-generated third harmonic must reach sufficient strength for detection. Second, propagation losses and coupling between multiple waveguide modes must be properly accounted for. Finally, the three-dimensional temporal trajectory traced by the optical field must retain a nonzero handedness when averaged over the full interaction volume. Here we show that all of these requirements can be satisfied simultaneously for a two-color field propagating through a capillary waveguide (Fig.~\ref{fig:setup}a). As a result, the electric field traces a spatially varying three-dimensional Lissajous figure with invariant handedness at every point along the propagation axis (Fig.~\ref{fig:setup}c). The two driving fields—a fundamental field $\vec{E}_1$ and its second harmonic $\vec{E}_2$—generate a third color at the sum frequency through three interfering nonlinear pathways (Fig.~\ref{fig:setup}b). Among these, one pathway (SFG) is chiral, whereas the other two (THG and FWM) are achiral.
\footnote{Chiral pathways are symmetry-forbidden in isotropic achiral media, and have opposite phase in chiral media of opposite handedness. Achiral pathways are allowed in both types of media and insensitive to their handedness.}

\begin{center}
    \centering
    \safeincludegraphics[width=0.99\linewidth]{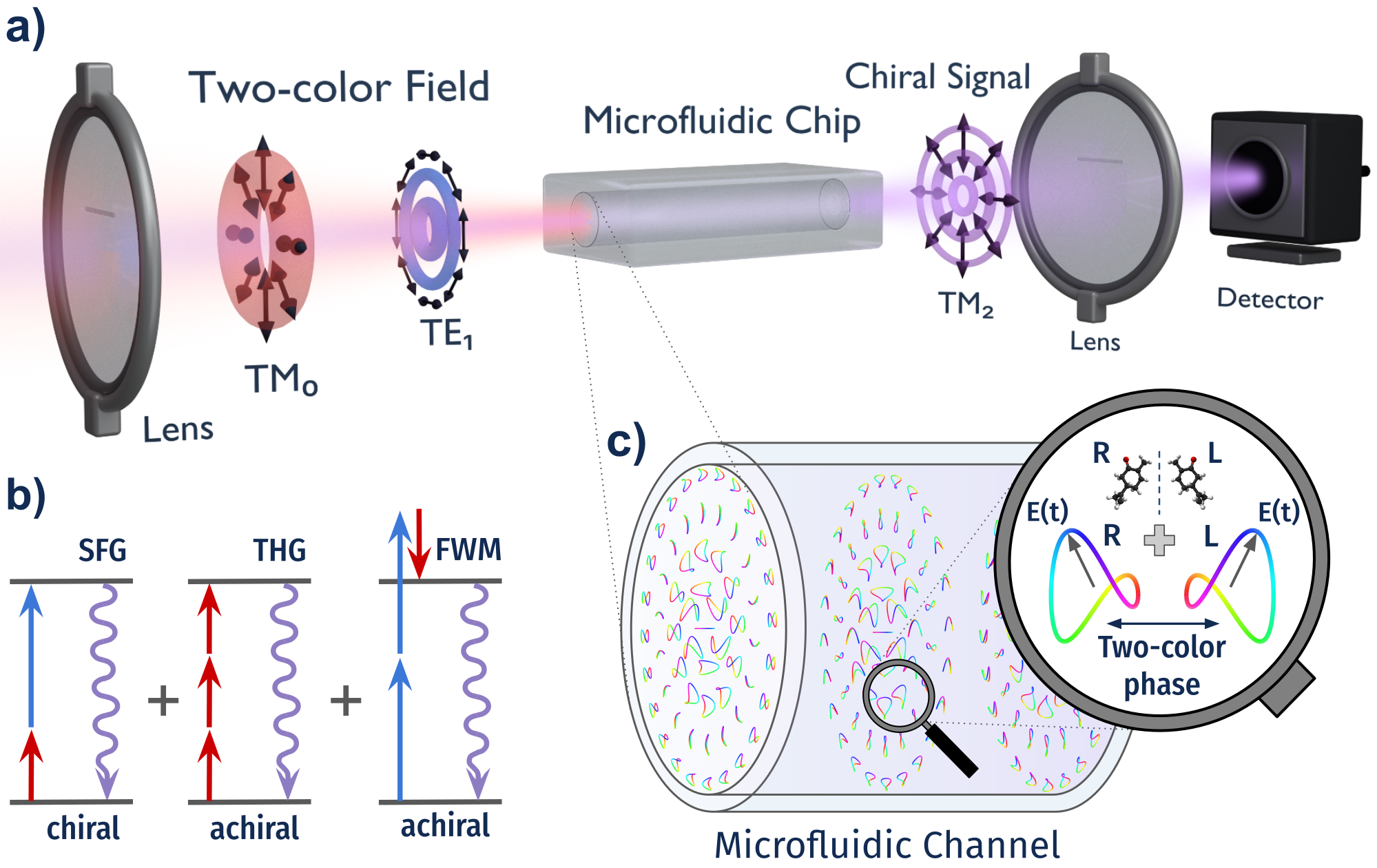}
    \captionof{figure}{(a) Proposed experimental scheme: at the waveguide input, the fundamental and its second harmonic are coupled into radially and azimuthally polarized modes. (b) In the waveguide, the third harmonic is generated through three competing chiral and achiral pathways. (c) Variation of the two-color phase results in a global change of the local synthetic chiral light.}
    \label{fig:setup}
\end{center}

To maximize the contribution of the chiral pathway, we engineer the modal composition of the driving fields. Circular hollow-core waveguides support a well-defined set of vector modes with known analytical form \cite{stratton2007electromagnetic} (see Methods~\ref{sec:methods_capillary_modes}).  We couple the fundamental field into the TM$_0$ mode, which provides the largest relative longitudinal field component, and the second harmonic into the TE$_1$ mode, which is strictly orthogonal at all times. The higher mode order of the second harmonic compensates dispersion between the fields, enabling near phase-matched propagation of the two drivers. This tailored modal combination simultaneously enhances the longitudinal driving field, preserves mutually orthogonal polarizations of all three colors, and optimizes the efficiency of chiral nonlinear interaction.

To model the macroscopic response we solve Maxwell's equations in the basis of eigenmodes $\vec{E}_\mu$ of an ideal capillary:
\begin{align}
    \vec{E}(\vec{r}, \omega) &=  \sum_\mu \vec{E}_\mu(\vec{r}_\perp, z,  \omega)\,,\\ \vec{E}_\mu(\vec{r}_\perp, z,  \omega)& =   c_\mu(z, \omega) \vec{F}_\mu(\vec{r}_\perp, \omega) e^{i \beta_\mu(\omega) z}\,.
\end{align}
 Here $\vec{F}_\mu$ describes the z-invariant shape of the three-dimensional eigenmode of an ideal capillary, $c_\mu$ is the complex amplitude of the constituent eigenmode and the modal propagation constant $\beta_\mu$ follows from the modal effective refractive index $N_\mu $:
\begin{align}
    \beta_\mu(\omega)=N_{\mu}(\omega)\frac{\omega}{c}\,.
    \end{align}
Using unidirectional pulse propagation approximation \cite{kolesik2004nonlinear} and taking into account that transverse field dominates over longitudinal we obtain the equation for the amplitude of a given mode:
\small{
\begin{align}
    \partial_z  c_\mu(z, \omega) = \frac{ i \omega}{2c N_{\mu}(\omega) \epsilon_0 } \frac{\int_A \tilde{\vec{P}}_{\mathrm{NL}}(\vec{r}_\perp, z, \omega) \cdot \vec{E}_{\mu }(\vec{r}_\perp, z, \omega)^* d\vec{r}_\perp }{\int_A | \vec{E}_\mu(\vec{r}_\perp, z,  \omega)|^2 d\vec{r}_\perp  }\,,
    \label{Propagation Upp}
\end{align}
}
where $\epsilon_0$ is the vacuum permittivity. The field-induced non-linear polarization $\vec{P}_{\mathrm{NL}}$ is at the heart of the propagation process. In the perturbative regime, the locally induced achiral and chiral nonlinear polarizations at the third-harmonic frequency are well established. The achiral contributions arise from two pathways,

\begin{align}
    \tilde{\vec{P}}_\text{ach}& (\vec{r}, 3\omega_0) \epsilon_0^{-1} = \nonumber \\ &\chi_\text{THG}^{(3)} \vec{E}_1(\vec{r}, \omega_0)^3 
    + \chi^{(3)}_\text{FWM}\vec{E}_2(\vec{r}, 2\omega_0)^2 \vec{E}_1^*(\vec{r}, \omega_0)\,,
    \label{achiral_response}
\end{align}

whereas the chiral contribution is given by
\begin{align}
    \tilde{\vec{P}}_\text{ch} (\vec{r}, 3\omega_0)  \epsilon_0^{-1}= \chi^{(2)}_\text{SFG} \bigg( \vec{E}_1(\vec{r}, \omega_0) \times \vec{E}_2(\vec{r},   2\omega_0) \bigg)\,.
    \label{chiral_response}
\end{align}
Here $\chi_\text{p}$ denotes the orientation-averaged nonlinear susceptibility associated with each nonlinear pathway \cite{Boyd, PhysRev.138.A1599, FISCHER2002463} (Fig.~\ref{fig:setup}b). Note that Eq.~\eqref{chiral_response} governs chiral SFG where enantio-sensitive signal is encoded in the direction of polarization $\vec{P}_\text{ch}$. In contrast, the enantio-sensitive response in our setup is encoded in the mode populations $|c_\mu(z, \omega)|^2$ in close analogy with the enantio-sensitive population transfer between electronic \cite{ordonez2024} or rotational \cite{perez, eibenberger, Lee2024} states.

The macroscopic nonlinear response is obtained by combining microscopic TDSE calculations of the nonlinear polarization with waveguide propagation (Methods~\ref{sec:methods_nonlinear_signal} and Methods~\ref{sec:methods_modal_propagation}). The induced polarization is projected onto the guided modes of the capillary, yielding the propagation-dependent modal amplitudes $c_\mu(z, 3\omega)$. Owing to the weak nonlinear conversion, mode–mode coupling is neglected and the total third-harmonic signal is obtained by summing the modal intensities,
\small{
\begin{align}
I(z,3\omega_0)\propto 
\sum_\mu 
\int_{3\omega_0-\Delta\omega/2}^{3\omega_0+\Delta\omega/2}
|c_\mu(z,\omega)|^2
\int_A |\vec{F}_\mu(\vec{r}_\perp,\omega)|^2
d\vec{r}_\perp\,d\omega \,,
\label{eq:intensity-summed-main}
\end{align}
}
where $\Delta \omega$ denotes the spectral bandwidth containing the third-harmonic signal.

\section{Phase-matched accumulation of enantio-sensitive signal}

Our system consists of a silica capillary filled with 50\,mbar of carvone gas at 404\,K with a length of up to 3 cm. We consider a widely used high-power laser platform: a Ti:sapphire-type driver pulse at 800\,nm, 30fs FWHM, 1.06 \textmu J and its second harmonic at 400\,nm, 30 fs FWHM, 0.59 \textmu J coupled into the 70\,\textmu m diameter capillary. The fundamental field is launched into the TM$_0$ mode of the capillary waveguide, while its second harmonic is launched into the TE$_1$ mode.

\begin{center}
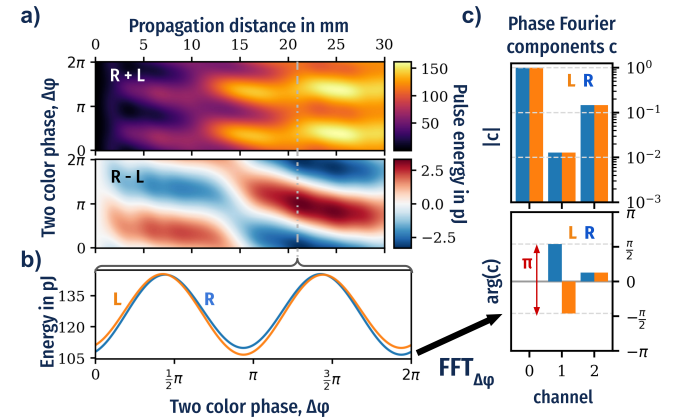

\safeincludegraphics[width=1.0\linewidth]{KeyResult.jpeg}
\captionof{figure}{Enantio-sensitive nonlinear signal in the capillary. (a) Achiral (top panel) and chiral (bottom panel) signal accumulation versus interaction length and two-color phase delay. (b) Enantiomer-resolved output versus phase delay at the propagation length of highest contrast. (c) Fourier/lock-in analysis isolates the background-free chiral component, channel 1; signals from opposite enantiomers are $\pi$ out of phase.}
\label{fig:results}
\end{center}

Figure~\ref{fig:results}a shows the accumulated chiral and achiral third-harmonic signal as a function of propagation distance $z$ and initial two-color phase delay. The total signal exhibits pronounced longitudinal beating superimposed on an overall growth. In both observables, the beating pattern remains smooth and regular, indicating that the response is carried predominantly by a single generated mode. The enantio-sensitive component displays a slower modulation reflecting its distinct phase-matching condition.
For a 2.1-cm-long capillary filled with dilute carvone gas (50 mbar), corresponding to only $\sim$120 pmol of molecules within the interaction volume, the enantio-sensitive output reaches the picojoule level while exhibiting a chiral dichroism $CD=2\frac{I_R-I_S}{I_R+I_S}$ of approximately 5\% (Figure~\ref{fig:results}b).
This raw chiral contrast already exceeds that of conventional linear chiroptical methods by orders of magnitude. As we show next, exploiting the phase symmetry of synthetic chiral light isolates a background-free chiral observable with the maximum possible dichroic contrast (Figure~\ref{fig:results}c).

This behavior is consistent with the modal dispersion of the capillary: the mode index $N$ of the third harmonic in the TM$_2$ mode lies very close to the modal indices set by the TM$_0$ (fundamental) and TE$_1$ (second-harmonic) drivers. In a perturbative picture (see Methods~\ref{sec:methods_perturbative_capillary}), the corresponding phase mismatches for the three pathways in Fig.~\ref{fig:setup}b take the form
\begin{eqnarray}
    &&\Delta k_{\mu}^\text{THG} = 3 \beta_{\text{TM}_0}(\omega_0) - \beta_\mu(3\omega_0), \label{eq:phasemismatchTHG} \\
    &&\Delta k_{\mu}^\text{FWM} = 2 \beta_{\text{TE}_1}(2\omega_0) - \beta_{\text{TM}_0}(\omega_0 )- \beta_\mu(3\omega_0),\label{eq:phasemismatchFWM} \\ 
    &&\Delta k_{\mu}^\text{SFG} = \beta_{\text{TM}_0}(\omega_0) + \beta_{\text{TE}_1}(2\omega_0 ) - \beta_\mu(3\omega_0).\label{eq:phasemismatchSHG}
\end{eqnarray}

Interference between these contributions produces the longitudinal beating in the generated intensity. A smaller phase mismatch increases the propagation length over which the nonlinear polarization adds constructively to the generated field. As a consequence of the modal-index proximity, the mode $\mu = \text{TM}_2$ minimizes the phase mismatch for all three nonlinear pathways, Eq.(\ref{eq:phasemismatchTHG}-\ref{eq:phasemismatchSHG}), and therefore dominates the nonlinear signal accumulation. 
The relevant coherence lengths extracted from the real parts of the mismatches are summarized in Table~\ref{tab:phasematching_lengths} and reveal the three characteristic propagation scales associated with THG, SFG, and FWM, respectively: approximately $\sim5$\,mm, $\sim8$\,mm and $\sim39$\,mm. Owing to its near-perfect phase matching, the FWM contribution accumulates uniformly with distance and therefore provides a slowly varying phase reference for interference with the other pathways. Notably the chiral pathway has a coherence length of $\sim8$\,mm, consistent with the dominant modulation scale observed in Fig.~\ref{fig:results}a. The same analysis accounts for the main features of the racemic signal, whose shorter modulation scale is associated with the $\sim8$\,mm coherence length of the THG pathway.

\begin{center}
    \begin{tabular}{c|c|c|c}
        $\text{L}_\text{co}$ &
        $\dfrac{\pi}{\text{Re}\big[ \Delta k_{\text{TM}_2}^\text{THG}\big]}$ &
        $\dfrac{\pi}{\text{Re}\big[ \Delta k_{\text{TM}_2}^\text{FWM}\big]}$ &
        $\dfrac{\pi}{\text{Re}\big[ \Delta k_{\text{TM}_2}^\text{SFG}\big]}$  \\
        \hline
        {[mm]} & 4.69 & 38.82 & 8.40 
    \end{tabular}
    \captionof{table}{Coherence lengths of the nonlinear pathways (Fig.~\ref{fig:setup}b). Here $\Delta k_\text{p}^\text{re}$ denotes the real part of the phase mismatch associated with pathway $p$.}
    \label{tab:phasematching_lengths}
\end{center}

\section{Chiral signal isolation via phase lock-in detection}

In the constellation presented here, the handedness of synthetic chiral light is determined by the relative phase between its two driving frequency components. A phase shift $ \phi'=\phi+\pi$ reverses the handedness of the driving field. Consequently, any chiral observable must change sign, whereas achiral observables remain invariant. Thus, chiral contributions are odd and achiral contributions are even under $\phi'=\phi+\pi$. This symmetry of the driving field can be exploited by recording the signal as a function of the two-color phase and decomposing it into its Fourier components. The odd Fourier harmonics isolate the chiral response, while the even harmonics contain only achiral contributions, enabling background-free extraction of the chiral signal.

Beyond separating chiral and achiral contributions, the Fourier decomposition uniquely resolves the underlying multiphoton pathways. The $n$-th Fourier harmonic identifies interference between pathways differing by $n$ 2$\omega$ photons. The Fourier components separate interferences of distinct multiphoton processes such that in combination with an independent measurement of THG intensity the molecule specific field amplitudes and relative nonlinear phases can be reconstructed, allowing simultaneous access to molecular handedness, concentration and specificity, see Methods \ref{sec:methods_Lock-in}. 

Figures~\ref{fig:results}b,c show the phase-dependent signal and Fourier/lock-in decomposition for both enantiomers in the 70\,\textmu m capillary at $z=21$\,mm. While the lock-in amplitudes are nearly identical for the two enantiomers, the odd Fourier components are shifted by $\pi$, providing an enantiomer-specific phase signature. The first lock-in harmonic therefore presents a robust, background-free enantio-sensitive observable over a range of propagation lengths with the maximum possible dichroic contrast of 200\%.

\section{The confinement advantage}
\label{sec:confinement_advantage}

To quantify the advantages of waveguide confinement, we compare enantio-sensitive nonlinear frequency conversion in a gas-filled capillary with an equivalent free-space focusing geometry (Figure~\ref{fig:focus_setup}b). To ensure a direct comparison, the input beams are chosen such that the focal fields reproduce the spatial profile, polarization, and peak intensity of the guided modes at the capillary entrance (Figure~\ref{fig:focus_setup}a). Both geometries use the same gas density, interaction length, and microscopic nonlinear response. The latter is obtained from pulse-averaged TDSE calculations within a perturbative response model, using the pulse, focusing, capillary, and medium parameters listed in Methods~\ref{sec:methods_perturbative_parameters}. Although this approximation does not reproduce the strongly driven resonant dynamics discussed above, it isolates the influence of propagation and modal confinement.

\begin{widetext}
\begin{center}
\safeincludegraphics[width=0.95\linewidth]{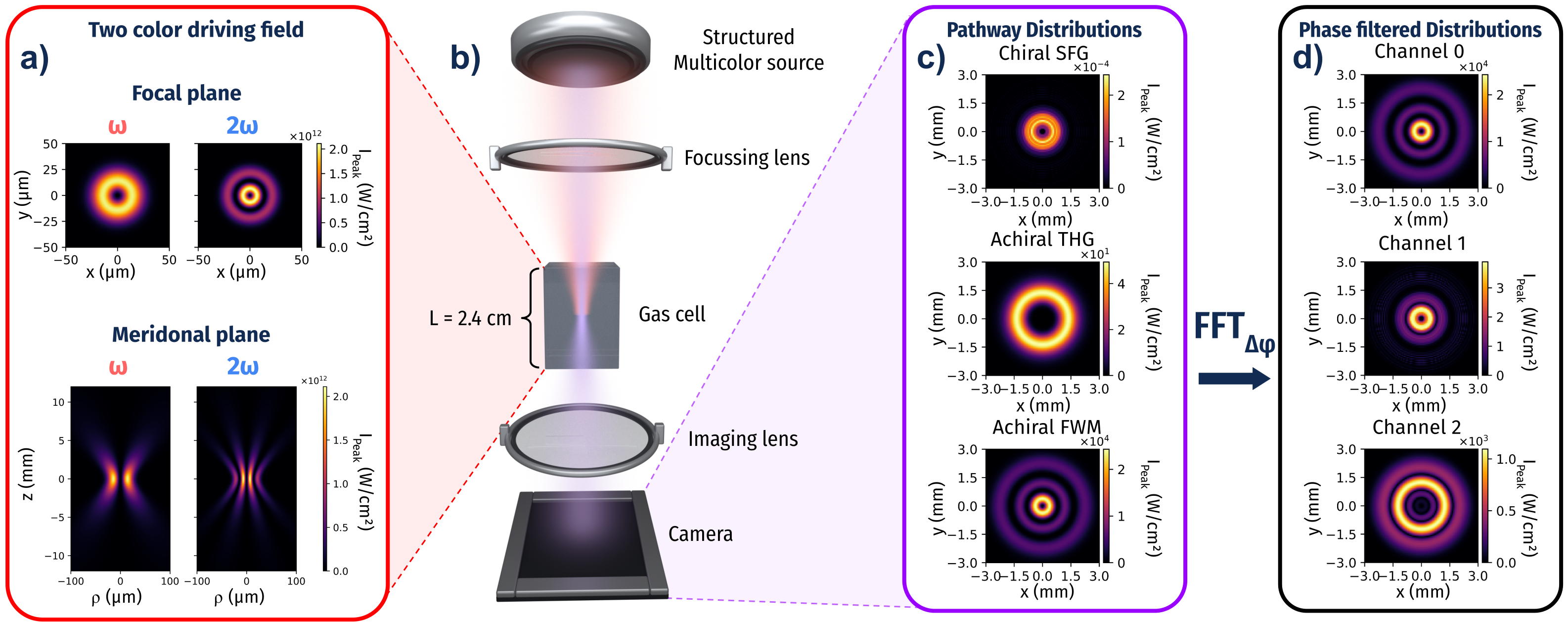}
\captionof{figure}{Equivalent free-space setup. (a) In the focal plane, the fields have D4$\sigma$ diameters matching the respective modes in a 70\,\textmu m capillary. (b) The structured multicolor field consists of collimated radially polarized LG$_{01}$ and azimuthally polarized LG$_{11}$ beams focused into a 2.4 cm long gas cell filled with carvone. The nonlinear signal generated by the individual nonlinear pathways is recollimated by an imaging lens and projected onto a camera. (c) Far-field intensity distributions generated by the recollimated individual nonlinear pathways. (d) Lock-in filtered, spatially resolved image signal resulting from pathway interferences.}
\label{fig:focus_setup}
\end{center}
\end{widetext}

Waveguide confinement enhances the nonlinear response through three complementary mechanisms. First, the guided modes maintain high intensity over the entire interaction length, whereas the free-space focus is typically localized to a much smaller length. Consequently, nonlinear polarization accumulates continuously throughout the capillary but only within a narrow region around the focal plane in free space.

Second, free-space focusing introduces rapidly varying Gouy and radial phase shifts that disrupt coherent signal buildup \cite{lastzka2007gouy,boydKleinman1968,olivier2008focusEngineeredTHG}. This is evident in Fig.~\ref{fig:comparison}a, where the SFG signal reaches its maximum before the focal plane and both SFG and THG accumulate only weakly. By contrast, the guided modes preserve well-defined longitudinal phase matching, allowing the nonlinear fields to build up coherently over centimeter-scale propagation (Fig.~\ref{fig:comparison}c). As a result, the SFG and THG signals are enhanced by one to two orders of magnitude. Four-wave-mixing (FWM) remains nearly phase matched in both geometries and therefore dominates the generated spectrum.

\begin{widetext}
\begin{center}
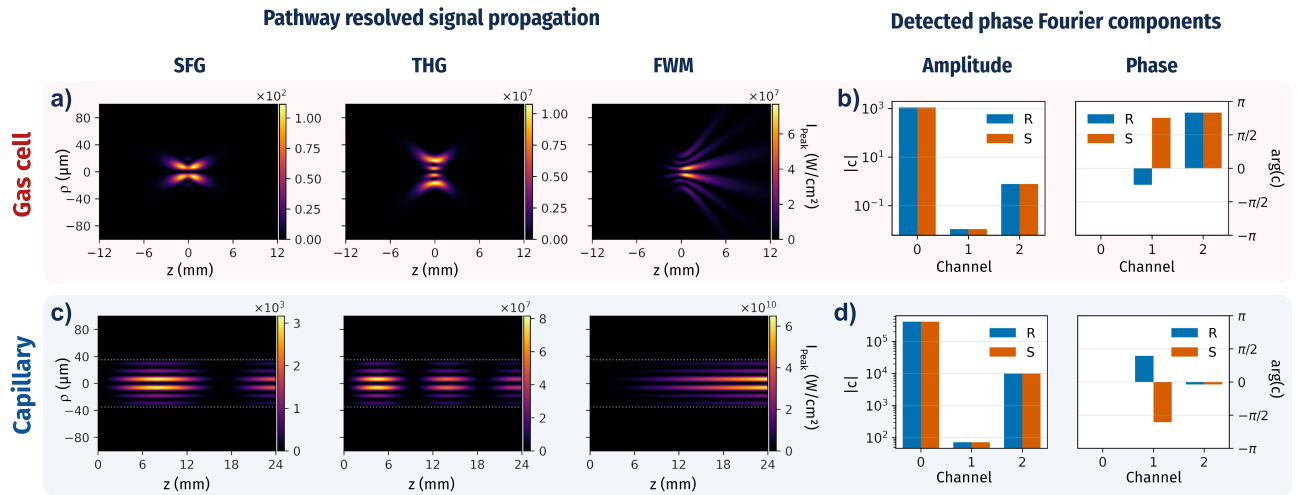

\safeincludegraphics[width=0.95\linewidth]{Comparison_cell_capillary.jpeg}
\captionof{figure}{(a) Pathway-resolved nonlinear signal accumulation in the 
meridional plane of the free-space setup. (b) Spatially integrated, phase-filtered signal far-field distributions. (c) Pathways-resolved nonlinear signal accumulation along the capillary. (d) Phase-filtered signal after capillary propagation.}
\label{fig:comparison}
\end{center}
\end{widetext}

Third, waveguide confinement naturally enforces spatial mode matching between the nonlinear pathways. In the free-space geometry, different nonlinear processes emit into distinct far-field distributions because of differences in phase matching, polarization, and transverse intensity profiles (Fig.~\ref{fig:focus_setup}c). Their interference is therefore limited by incomplete spatial overlap, reducing the efficiency of phase-sensitive detection (Fig.~\ref{fig:focus_setup}d). In the capillary, by contrast, all efficiently phase-matched pathways are emitted into the same guided mode, maximizing their overlap throughout the nonlinear interaction and any consequent measurement.

Finally, we compare the experimentally accessible observables by spatially integrating the phase-filtered far-field distributions in free space and comparing them with the phase-filtered capillary output (Fig.~\ref{fig:comparison}b,d). Both geometries enable background-free extraction of the enantio-sensitive signal. However, for identical input pulse energies, waveguide confinement enhances the detected signal by approximately three orders of magnitude. This improvement arises not from stronger microscopic nonlinearities, but from the combined effects of extended interaction length, improved phase matching, and intrinsic modal overlap.

\section{Outlook}

The guided architecture presented here establishes synthetic chiral light as a practical platform for nonlinear chiral spectroscopy. By combining modal engineering with phase-matched nonlinear frequency conversion, hollow dielectric waveguides enable coherent accumulation of enantio-sensitive signals over centimeter-scale distances, enhancing the detected response by orders of magnitude while preserving background-free discrimination through symmetry-selective Fourier analysis. 
Beyond enabling efficient enantio-sensitive detection, phase-resolved Fourier analysis provides access to a new class of molecular observables. By resolving the amplitudes and relative quantum phases of distinct multiphoton excitation pathways, it enables their experimental reconstruction. These quantities provide molecular fingerprints that originate entirely from ultrafast non-equilibrium electron dynamics, complementing conventional chiroptical and spectroscopic observables that probe equilibrium molecular structure.

More broadly, this work establishes the first photonic architecture specifically designed to generate, guide and exploit synthetic chiral light. This opens a new direction in which photonic structures become an active tool for engineering chiral light-matter interactions rather than merely delivering the optical field. The compatibility of hollow waveguides with mature microfluidic and lab-on-chip technologies \cite{monat2007integrated} provides a natural route toward compact integrated platforms for highly efficient chiral sensing.
Looking forward, photonic confinement offers opportunities that extend well beyond the present demonstration. Tailored waveguide geometries, integrated photonic circuits and metastructured optical environments \cite{deng2024advances,torresCubillo2024optimized} can be designed to generate chiral optical fields with engineered spatial, temporal and topological properties \cite{lu2014topological}. In particular, guided photonic structures provide a natural platform for realizing chiral topological light and exploring topologically robust enantio-sensitive observables that remain resilient against optical imperfections \cite{mayer2024chiralTopologicalLight, mayer2025chiralExceptionalPoints}.

An important next step will be extending coherent signal accumulation from dilute gases to condensed media. Developing quasi-phase-matching strategies for synthetic chiral light will enable efficient nonlinear chiral spectroscopy in liquids, paving the way toward integrated optofluidic devices capable of real-time analysis of chemically and biologically relevant samples.

\section{Methods}
\label{sec:methods}

\subsection{Capillary modes}
\label{sec:methods_capillary_modes}

Any electromagnetic field $\vec{E}$ inside the waveguide can be decomposed into eigenmodes of a circular capillary:

\begin{align}
    \vec{E}(\vec{r}, \omega) = \sum_\mu c_\mu(z, \omega) \vec{F}_\mu(\vec{r}_\perp, \omega) e^{i \beta_\mu(\omega) z},\\
    \vec{H}(\vec{r},\omega) = \sum_\mu c_\mu(z, \omega) \vec{G}_\mu(\vec{r}_\perp, \omega) e^{i \beta_\mu(\omega) z} .
\end{align}
 Here $\vec{F}_\mu$ and $\vec{G}_\mu$ describe the $z$-invariant shape of the three-dimensional electromagnetic eigenmode, $c_\mu$ is the complex amplitude of the constituent eigenmode and the modal propagation constant $\beta_\mu$ follows from the modal effective refractive index $N_\mu $:\\

\begin{align}
    \beta_\mu(\omega)=N_{\mu}(\omega)\frac{\omega}{c}\,.
\end{align}

We choose the modes of the driving fields with the goal of maximizing the efficiency of the chiral pathway. Our choice of cylindrically symmetric TE and TM modes maximizes the available longitudinal component of one field while enforcing the other to be strictly perpendicular. 
The natural vector modes of circular hollow-core waveguides can be obtained from the known general solution of Maxwell's equations in a cylindrical geometry\cite{stratton2007electromagnetic}, and we can write them in analytical form:

Let $a$ be the capillary inner radius and $n_{co}$, $n_{cl}$ the refractive index of the core and cladding material respectively. The electric fields of TE modes have strictly azimuthal components. TM modes, on the other hand, lack any azimuthal components but show both radial and longitudinal components:

\begin{eqnarray}
    &&\vec{F}^{\text{TE}_m}(\rho  ,  \omega) = J_1  \bigg( \frac{\omega}{c}  \sqrt{n^2_{co}(\omega)  - N_{\text{TE}_m}^2(\omega)} \rho  \bigg) \hat{\phi} \,, \\
    &&\vec{F}^{\text{TM}_m}(\rho  , \omega) = i
     J_1\bigg( \frac{\omega}{c} \sqrt{n^2_{co}(\omega)  - N_{\text{TM}_m}^2(\omega)} \rho  \bigg)  \hat{\rho} + \nonumber \\
      &&\sqrt{\frac{n^2_{co}(\omega)}{N_{\text{TM}_m}^2(\omega)} - 1} J_0 \bigg( \frac{\omega}{c} \sqrt{n^2_{co}(\omega)  - N_{\text{TM}_m}^2(\omega)} \rho  \bigg) \hat{z}\,,
     \label{NormTMmodes}
\end{eqnarray}

where the effective mode indices $N_{\text{TE}_m}(\omega)$ of TE modes and $N_{\text{TM}_m}(\omega)$ of TM modes have to be determined numerically from the individual characteristic equations:

{\small
\begin{equation}
\begin{aligned}
\frac{J_1\!\left( \frac{\omega a}{c}
\sqrt{n^2_{co}(\omega) - N_{\text{TE}_m}^2(\omega)} \right)}
     {J_0\!\left( \frac{\omega a}{c}
\sqrt{n^2_{co}(\omega) - N_{\text{TE}_m}^2(\omega)} \right)}
=
\frac{\sqrt{n^2_{co}(\omega) - N_{\text{TE}_m}^2(\omega)}}
     {\sqrt{n^2_{cl}(\omega) - N_{\text{TE}_m}^2(\omega)}} & \\
\quad{}\times
\frac{H_1^{(1)}\!\left( \frac{\omega a}{c}
\sqrt{n^2_{cl}(\omega) - N_{\text{TE}_m}^2(\omega)} \right)}
     {H_0^{(1)}\!\left( \frac{\omega a}{c}
\sqrt{n^2_{cl}(\omega) - N_{\text{TE}_m}^2(\omega)} \right)} & \,,
\end{aligned}
\end{equation}
}

for TE modes, and

{\small
\begin{equation}
\begin{aligned}
\frac{J_1\!\left( \frac{\omega a}{c}
\sqrt{n^2_{co}(\omega) - N_{\text{TM}_m}^2(\omega)} \right)}
     {J_0\!\left( \frac{\omega a}{c}
\sqrt{n^2_{co}(\omega) - N_{\text{TM}_m}^2(\omega)} \right)}
=
\frac{\sqrt{n^2_{co}(\omega) - N_{\text{TM}_m}^2(\omega)}}
     {\sqrt{n^2_{cl}(\omega) - N_{\text{TM}_m}^2(\omega)}}&  \\
\quad{}\times \frac{n_{cl}^2(\omega)}{n_{co}^2(\omega)}
\frac{H_1^{(1)}\!\left( \frac{\omega a}{c}
\sqrt{n^2_{cl}(\omega) - N_{\text{TM}_m}^2(\omega)} \right)}
     {H_0^{(1)}\!\left( \frac{\omega a}{c}
\sqrt{n^2_{cl}(\omega) - N_{\text{TM}_m}^2(\omega)} \right)}& \,,
\end{aligned}
\end{equation}
}
for TM modes. \\

Analogous field solutions and characteristic equations exist for HE and EH modes, but are not needed in the scope of this work.

\subsection{Macroscopic nonlinear propagation equation}
\label{sec:methods_modal_propagation}

The frequency-domain Maxwell equations in SI units are
\begin{align}
    \vec{\nabla}\times
    \tilde{\vec{E}}(\rho,\phi,z,\omega)
    &=
    i\omega\mu_0
    \tilde{\vec{H}}(\rho,\phi,z,\omega),
    \label{eq:maxwell_E}
    \\
    \vec{\nabla}\times
    \tilde{\vec{H}}(\rho,\phi,z,\omega)
    &=
    -i\omega\epsilon_0
    n^2(\rho,\omega)
    \tilde{\vec{E}}(\rho,\phi,z,\omega)
    \nonumber\\
    &\quad
    -
    i\omega
    \tilde{\vec{P}}_{\mathrm{NL}}
    (\rho,\phi,z,\omega).
    \label{eq:maxwell_H}
\end{align}
The nonlinear polarization is treated as a source term. 

Consistent with the mode expansion introduced above, the right electromagnetic eigenmodes of the capillary are written as
\begin{align}
    \vec{E}^{R,\mu}(\rho,\phi,z,\omega)
    =
    \vec{F}_{\mu}(\rho,\phi,\omega)
    e^{i\beta_\mu(\omega)z},\\
    \vec{H}^{R,\mu}(\rho,\phi,z,\omega)
    =
    \vec{G}_{\mu}(\rho,\phi,\omega)
    e^{i\beta_\mu(\omega)z}.
\end{align}
For lossy, leaky, or otherwise non-Hermitian waveguides, the usual Hermitian mode product is in general not the correct orthogonality relation. Instead, the modal projection must be written in terms of the left and right Maxwell modes, or equivalently in terms of the Lorentz reciprocity product \cite{Xu2015,Chen2019}. We therefore introduce the corresponding left, or adjoint, modes as
\begin{align}
    \vec{E}^{L,\nu}(\rho,\phi,z,\omega)
    &=
    \vec{F}^{L}_{\nu}(\rho,\phi,\omega)
    e^{-i\beta_\nu(\omega)z},
    \\
    \vec{H}^{L,\nu}(\rho,\phi,z,\omega)
    &=
    \vec{G}^{L}_{\nu}(\rho,\phi,\omega)
    e^{-i\beta_\nu(\omega)z}.
\end{align}
The right and left eigenmodes are normalized by
\begin{align}
    \mathcal N_{\nu\mu}(\omega)
    &=
    \int_{A_{\mathrm{reg}}}
    \left[
        \vec{F}^{L}_{\nu}
        \times
        \vec{G}_{\mu}
        -
        \vec{F}_{\mu}
        \times
        \vec{G}^{L}_{\nu}
    \right]
    \cdot
    \vec{e}_z
    \,dA
    \nonumber\\
    &=
    \mathcal N_\mu(\omega)\delta_{\mu\nu}.
    \label{eq:lorentz_biorthogonality}
\end{align}
Here
\begin{align}
    \mathcal N_\mu(\omega)
    =
    \int_{A_{\mathrm{reg}}}
    \left[
        \vec{F}^{L}_{\mu}
        \times
        \vec{G}_{\mu}
        -
        \vec{F}_{\mu}
        \times
        \vec{G}^{L}_{\mu}
    \right]
    \cdot
    \vec{e}_z
    \,dA
    \label{eq:lorentz_norm}
\end{align}
is the Lorentz-reciprocity normalization. The domain \(A_{\mathrm{reg}}\) denotes the transverse domain on which the modes are normalized. For bound modes this can be the full transverse plane. For exact leaky modes, the outgoing fields are not square-integrable on \(\mathbb R^2\), and \(A_{\mathrm{reg}}\) must instead be interpreted as a regularized domain, for example a complex-stretched or PML-normalized computational domain \cite{SammutSnyder1976,Sauvan2022}.

We expand the nonlinear electric and magnetic fields in the right eigenmode basis:
\begin{align}
    \tilde{\vec{E}}(\rho,\phi,z,\omega)
    &=
    \sum_\mu
    c_\mu(z,\omega)
    \vec{F}_{\mu}(\rho,\phi,\omega)
    e^{i\beta_\mu(\omega)z},
    \label{eq:E_expansion}
    \\
    \tilde{\vec{H}}(\rho,\phi,z,\omega)
    &=
    \sum_\mu
    c_\mu(z,\omega)
    \vec{G}_{\mu}(\rho,\phi,\omega)
    e^{i\beta_\mu(\omega)z}.
    \label{eq:H_expansion}
\end{align}
The coefficients \(c_\mu(z,\omega)\) describe the nonlinear evolution, while the factor \(e^{i\beta_\mu z}\) contains the linear phase and attenuation.

The propagation equation is obtained from the Lorentz reciprocity relation between the nonlinear field \((\tilde{\vec{E}},\tilde{\vec{H}})\) and an adjoint source-free mode \((\vec{E}^{L,\nu},\vec{H}^{L,\nu})\). Starting from the vector identity
\begin{align}
    &\vec{\nabla}\cdot
    \left(
        \vec{E}^{L,\nu}
        \times
        \tilde{\vec{H}}
        -
        \tilde{\vec{E}}
        \times
        \vec{H}^{L,\nu}
    \right)
    \nonumber\\
    &=
    \tilde{\vec{H}}
    \cdot
    \left(
        \vec{\nabla}\times
        \vec{E}^{L,\nu}
    \right)
    -
    \vec{E}^{L,\nu}
    \cdot
    \left(
        \vec{\nabla}\times
        \tilde{\vec{H}}
    \right)
    \nonumber\\
    &\quad
    -
    \vec{H}^{L,\nu}
    \cdot
    \left(
        \vec{\nabla}\times
        \tilde{\vec{E}}
    \right)
    +
    \tilde{\vec{E}}
    \cdot
    \left(
        \vec{\nabla}\times
        \vec{H}^{L,\nu}
    \right),
    \label{eq:vector_identity}
\end{align}
we insert Eqs.~\eqref{eq:maxwell_E} and \eqref{eq:maxwell_H}, together with the homogeneous adjoint-mode equations
\begin{align}
    \vec{\nabla}\times
    \vec{E}^{L,\nu}
    &=
    i\omega\mu_0
    \vec{H}^{L,\nu},
    \\
    \vec{\nabla}\times
    \vec{H}^{L,\nu}
    &=
    -i\omega\epsilon_0
    n^2(\rho,\omega)
    \vec{E}^{L,\nu}.
\end{align}
The linear terms cancel pairwise, and the only remaining contribution is the nonlinear source term. This gives the local Lorentz reciprocity relation
\begin{align}
    &\vec{\nabla}\cdot
    \left(
        \vec{E}^{L,\nu}
        \times
        \tilde{\vec{H}}
        -
        \tilde{\vec{E}}
        \times
        \vec{H}^{L,\nu}
    \right)
    \nonumber\\
    &\quad=
    i\omega
    \vec{E}^{L,\nu}
    \cdot
    \tilde{\vec{P}}_{\mathrm{NL}} .
    \label{eq:local_reciprocity}
\end{align}
This is the unconjugated Lorentz reciprocity relation used for the modal projection. It is exact for reciprocal media when the actual and adjoint problems share the same linear material distribution \cite{Sukhorukov2014}.

Integrating Eq.~\eqref{eq:local_reciprocity} over the regularized transverse domain gives
\begin{align}
    &\partial_z
    \int_{A_{\mathrm{reg}}}
    \left[
        \vec{E}^{L,\nu}
        \times
        \tilde{\vec{H}}
        -
        \tilde{\vec{E}}
        \times
        \vec{H}^{L,\nu}
    \right]
    \cdot
    \vec{e}_z
    \,dA
    \nonumber\\
    &\quad=
    i\omega
    \int_{A_{\mathrm{NL}}}
    \vec{E}^{L,\nu}
    \cdot
    \tilde{\vec{P}}_{\mathrm{NL}}
    \,dA .
    \label{eq:integrated_reciprocity}
\end{align}
The right-hand side only has support in the nonlinear region \(A_{\mathrm{NL}}\), which for the present hollow waveguide is the gas-filled core.

Substitution of Eqs.~\eqref{eq:E_expansion} and \eqref{eq:H_expansion} into the overlap integral on the left-hand side gives
\begin{align}
    &
    \int_{A_{\mathrm{reg}}}
    \left[
        \vec{E}^{L,\nu}
        \times
        \tilde{\vec{H}}
        -
        \tilde{\vec{E}}
        \times
        \vec{H}^{L,\nu}
    \right]
    \cdot
    \vec{e}_z
    \,dA
    \nonumber\\
    &=
    \sum_\mu
    c_\mu(z,\omega)
    e^{i[\beta_\mu(\omega)-\beta_\nu(\omega)]z}
    \nonumber\\
    &\quad\times
    \int_{A_{\mathrm{reg}}}
    \left[
        \vec{F}^{L}_{\nu}
        \times
        \vec{G}_{\mu}
        -
        \vec{F}_{\mu}
        \times
        \vec{G}^{L}_{\nu}
    \right]
    \cdot
    \vec{e}_z
    \,dA .
\end{align}
Using Eq.~\eqref{eq:lorentz_biorthogonality}, only the term \(\mu=\nu\) remains:
\small{
\begin{align}
    &
    \int_{A_{\mathrm{reg}}}
    \left[
        \vec{E}^{L,\nu}
        \times
        \tilde{\vec{H}}
        -
        \tilde{\vec{E}}
        \times
        \vec{H}^{L,\nu}
    \right]
    \cdot
    \vec{e}_z
    \,dA
    \nonumber\\
    &\quad=
    c_\nu(z,\omega)\mathcal N_\nu(\omega).
    \label{eq:projected_overlap}
\end{align}
}
Since the waveguide is longitudinally invariant, \(\mathcal N_\nu\) is independent of \(z\). Hence Eq.~\eqref{eq:integrated_reciprocity} becomes
\begin{align}
    &\mathcal N_\nu(\omega)
    \partial_z c_\nu(z,\omega)
    \nonumber\\
    &\quad=
    i\omega
    \int_{A_{\mathrm{NL}}}
    \vec{F}^{L}_{\nu}(\rho,\phi,\omega)
    e^{-i\beta_\nu(\omega)z}
    \cdot
    \tilde{\vec{P}}_{\mathrm{NL}}
    (\rho,\phi,z,\omega)
    \,dA .
\end{align}
The rigorous modal propagation equation is therefore
\begin{align}
    &\partial_z c_\nu(z,\omega)
    \nonumber\\
    &\quad=
    \frac{i\omega}{\mathcal N_\nu(\omega)}
    \int_{A_{\mathrm{NL}}}
    \vec{F}^{L}_{\nu}(\rho,\phi,\omega)
    e^{-i\beta_\nu(\omega)z}
    \cdot
    \tilde{\vec{P}}_{\mathrm{NL}}
    (\rho,\phi,z,\omega)
    \,dA .
    \label{eq:rigorous_lorentz_mode_equation}
\end{align}
If the modes are exact leaky modes, the normalization in Eq.~\eqref{eq:lorentz_norm} is only valid in the regularized sense described above; it is not an ordinary integral over \(\mathbb R^2\).

For the hollow waveguide capillary considered here, the diameter is 70 \textmu m, thus the core radius $a= $ 35 \textmu m. At the optical and NIR frequencies used here,
\begin{align}
    k_0a
    =
    \frac{\omega}{c}a
    \gg 1 .
\end{align}
For low-order modes, the longitudinal components are therefore small,
\begin{align}
    \frac{|E_z|}{|E_\perp|}
    \sim
    \mathcal O
    \left(
        \frac{1}{k_0a}
    \right),
    \qquad
    \frac{|H_z|}{|H_\perp|}
    \sim
    \mathcal O
    \left(
        \frac{1}{k_0a}
    \right).
\end{align}
In the present application the nonlinear source is localized inside the hollow core, and the propagation length is only a few centimeters. We therefore use the low-order modes as weakly leaky, restricted-domain modes over the physical region relevant to the nonlinear interaction. This is an approximation: exact leaky modes are not proper power-orthogonal modes on the full open transverse plane, but weakly leaky modes may be treated as approximately power-orthogonal within a restricted spatial region as shown by Sammut et al. \cite{SammutSnyder1976}. In this restricted-domain limit, the left mode approaches the reciprocal conjugate of the right mode,
\begin{align}
    \vec{F}^{L}_{\nu}
    \approx
    \vec{F}_{\nu}^{*},
    \qquad
    \vec{G}^{L}_{\nu}
    \approx
    -
    \vec{G}_{\nu}^{*}.
    \label{eq:weak_nonhermitian_adjoint}
\end{align}
Using this approximation in Eq.~\eqref{eq:lorentz_norm}, with the normalization now taken over the physical region \(A_{\mathrm{core}}\), gives
\begin{align}
    \mathcal N_\nu(\omega)
    &\approx
    \int_{A_{\mathrm{core}}}
    \left[
        \vec{F}_{\nu}^{*}
        \times
        \vec{G}_{\nu}
        +
        \vec{F}_{\nu}
        \times
        \vec{G}_{\nu}^{*}
    \right]
    \cdot
    \vec{e}_z
    \,dA
    \nonumber\\
    &=
    2
    \mathrm{Re}
    \int_{A_{\mathrm{core}}}
    \left[
        \vec{F}_{\nu}
        \times
        \vec{G}_{\nu}^{*}
    \right]
    \cdot
    \vec{e}_z
    \,dA .
\end{align}
The corresponding restricted-domain modal power is
\begin{align}
    P_\nu(\omega)
    =
    \frac{1}{2}
    \mathrm{Re}
    \int_{A_{\mathrm{core}}}
    \left[
        \vec{F}_{\nu}
        \times
        \vec{G}_{\nu}^{*}
    \right]
    \cdot
    \vec{e}_z
    \,dA .
\end{align}
Therefore,
\begin{align}
    \mathcal N_\nu(\omega)
    \approx
    4P_\nu(\omega).
    \label{eq:N_approx_4P}
\end{align}
With this weakly leaky normalization approximation, Eq.~\eqref{eq:rigorous_lorentz_mode_equation} reduces to
\begin{align}
    &\partial_z c_\nu(z,\omega)
    \nonumber\\
    &\approx
    \frac{i\omega}{4P_\nu(\omega)}
    \int_{A_{\mathrm{core}}}
    \vec{F}_{\nu}^{*}(\rho,\phi,\omega)
    e^{-i\beta_\nu(\omega)z}
    \cdot
    \tilde{\vec{P}}_{\mathrm{NL}}
    (\rho,\phi,z,\omega)
    \,dA .
    \label{eq:power_normalized_equation}
\end{align}
This expression keeps the full vector mode profile and is the weakly leaky, restricted-domain approximation of the Lorentz-reciprocity result.

Using the large-core impedance approximation, the normalization can be written in terms of the electric mode profile alone. Since the same low-order modes already satisfy \(k_0a\gg1\) and have small longitudinal components, their transverse fields can be treated locally as a forward-propagating waveguide field with
\begin{align}
    \vec{G}_{\nu}
    \approx
    \frac{1}{Z_\nu}
    \vec{e}_z
    \times
    \vec{F}_{\nu},
    \qquad
    Z_\nu
    \approx
    \frac{Z_0}{N_\nu},
    \label{eq:impedance_approximation}
\end{align}
where \(Z_0=\mu_0 c\) and \(Z_\nu\) are the vacuum and modal impedance, respectively. This is the usual transverse-field or large-core approximation for weakly guiding waveguide modes \cite{SnyderLove1983}. It gives
\begin{align}
    \left[
        \vec{F}_{\nu}
        \times
        \vec{G}_{\nu}^{*}
    \right]
    \cdot
    \vec{e}_z
    \approx
    \frac{N_\nu(\omega)}{Z_0}
    \left|
        \vec{F}_{\nu}
    \right|^2 .
\end{align}
Using \(1/Z_0=\epsilon_0 c\), this yields
\begin{align}
    P_\nu(\omega)
    &\approx
    \frac{N_\nu(\omega)\epsilon_0 c}{2}
    \int_{A_{\mathrm{core}}}
    \left|
        \vec{F}_{\nu}(\rho,\phi,\omega)
    \right|^2
    \,dA,
    \\
    \mathcal N_\nu(\omega)
    &\approx
    2N_\nu(\omega)\epsilon_0 c
    \int_{A_{\mathrm{core}}}
    \left|
        \vec{F}_{\nu}(\rho,\phi,\omega)
    \right|^2
    \,dA .
    \label{eq:electric_field_norm_approx}
\end{align}
Substitution into Eq.~\eqref{eq:power_normalized_equation} yields the electric-field-normalized 
\begin{align}
    &\partial_z c_\nu(z,\omega)
    \approx
    i
    \frac{\omega}{2cN_\nu(\omega)\epsilon_0}\nonumber\\
    &\quad
    \frac{
        \displaystyle
        \int_{A_{\mathrm{core}}}
        \vec{F}_{\nu}^{*}(\rho,\phi,\omega)
        e^{-i\beta_\nu(\omega)z}
        \cdot
        \tilde{\vec{P}}_{\mathrm{NL}}
        (\rho,\phi,z,\omega)
        \,dA
    }{
        \displaystyle
        \int_{A_{\mathrm{core}}}
        \left|
            \vec{F}_{\nu}(\rho,\phi,\omega)
        \right|^2
        \,dA
    } .
    \label{eq:electric_field_normalized_equation}
\end{align}
This last equation additionally assumes that the low-order hollow-waveguide modes are weakly leaky, mostly transverse, and well described by the local impedance relation in Eq.~\eqref{eq:impedance_approximation}.

\subsection{Computation of the nonlinear polarization and signal accumulation}
\label{sec:methods_nonlinear_signal}

In our numerical modeling, we numerically integrate Eq.~\eqref{eq:electric_field_normalized_equation} to compute the accumulated nonlinear signal amplitude.
The main computational hurdle lies in obtaining the induced nonlinear polarization $\vec{P}_\text{NL}$ in randomly oriented chiral molecules. The conditions we present here, i.e., strong fields and resonant driving, make purely perturbative estimations of the nonlinear susceptibilities unreliable. 
We therefore calculate the induced polarization for many molecular orientations separately to obtain the time-dependent induced polarization at a single point by numerically solving the time-dependent Schr\"{o}dinger equation (TDSE) in the basis set of the field-free eigenstates using an explicit Runge-Kutta method of order 8 in the presence of the optical fields.\\

The molecular structure of the lowest-energy conformer of (R)-carvone was adopted from Ref. \citenum{lambert_optical_2012} and subsequently reoptimized to ensure consistency using Density Functional Theory (DFT) \cite{kohn1996dftElectronicStructure,ziegler1991approximateDft}. The calculations employed the B3LYP exchange–correlation functional \cite{becke_density-functional_1993,lee_development_1988,becke_density-functional_1988} together with the 6-311G(d,p) basis set \cite{krishnan1980basis,frisch1984supplementary}, as implemented in Gaussian 16 \cite{g16}. Excitation energies and transition dipole moments for the lowest 100 excited states were obtained using time-dependent DFT at the CAM-B3LYP/d-aug-cc-pVDZ level \cite{CAMB3LYP,augccpvdz_1,augccpvdz_2,augccpvdz_3}. Transition dipole moments between excited states were further evaluated using the Multiwfn program \cite{Multiwfn}.

To simulate a randomly oriented ensemble, the TDSE simulations were run for 570 molecular orientations and the averaged induced polarization was obtained using equation
\begin{equation}
    \boldsymbol{P} = \frac{1}{8\pi^{2}}\int_{0}^{2\pi}\int_{0}^{2\pi}\int_{0}^{\pi}\boldsymbol{P}_{\chi\phi\theta}\sin(\theta)d\chi d\phi d\theta   \,,
\end{equation}
where $\chi, \phi, \theta$ are the Euler angles and $\boldsymbol{P}_{\chi\phi\theta}$ is the induced polarization in the laboratory frame for a given molecular orientation.
The integral over $\phi$, $\theta$ was approximated using the Lebedev quadrature \cite{Lebedev1999AQF} of order 9, which gives 38 points on a sphere ($\phi$, $\theta$).
Integration over $\chi$ was performed using the trapezoid method including 15 rotations around each Lebedev axis.\\

As we are interested in the nonlinear signal accumulated in the capillary for different initial two-color input phases, calculating the pulse propagation for a single input two-color phase is computationally demanding because many points are needed to describe the field geometry. 
By using strictly cylindrical modes we reduce the computational complexity significantly as we only have to calculate dynamics along the radius of the capillary.
We exploit another property of our system to make our computations even more efficient;
Over the propagation scales considered in this work, the group delay accumulated between individual colors is less than 5\% of the driving pulse duration (FWHM), and the driving-mode power loss is below 10\%.

We therefore make the crucial approximation to treat the propagating driving field as a lossless, non-depleting two-color CW field with a shared envelope throughout the propagation process. We assume that the fiber dispersion only causes the two colors under the envelope to acquire a relative phase delay as they propagate.\\

These assumptions reduce the full $3D+1$ ($\rho$, $z$, $t$, $\Delta\phi_0$) computational problem to the calculation of the induced polarization of the field at a given two-color phase along the radius.

\begin{eqnarray}
    \tilde{\vec{P}}_{\mathrm{NL}}(\rho, \phi, z, \omega, \Delta\phi_0)  \approx \tilde{\vec{P}}_{\mathrm{NL}}(\rho, z(\Delta \phi), \omega, \Delta \phi_0)
    \label{Polarization_comp}
\end{eqnarray}

where the mapping of propagation distance to two-color phase follows the relative phase accumulation of the two driving fields:

\begin{eqnarray}
    (\beta_{\text{TE}_1}(2\omega_0)- 2 \beta_{\text{TM}_0}(\omega_0))z = \Delta \phi \,.
    \label{phase_z_map}
\end{eqnarray}

With this approximation we describe the entire nonlinear polarization in the capillary by only calculating the induced polarization along the radius for a number of two-color phase delays, which can be used in a straightforward fashion upon change of the two-color input phase.\\

The resulting nonlinear polarization distribution is projected onto a truncated set of waveguide modes (here the first 11 TM modes), yielding a set of generated-mode amplitudes as a function of $z$.
Since the generated signals are weak, their influence on each other is neglected, allowing us to simply add up their contributions. The total signal intensity at a given distance in the capillary is computed by: 

\small{
\begin{align}
    \text{I}(z,3\omega_0) \sim
    \sum_\mu \int_{3\omega_0 - \frac{\Delta \omega}{2}}^{3\omega_0 + \frac{\Delta \omega}{2}} | c_\mu(z, \omega) |^2 \int_A|\vec{F}_\mu(\vec{r}_\perp, \omega ) |^2 d\vec{r}_\perp d\omega \,.
    \label{eq:intensity-summed-methods}
\end{align}
}
The modal population of a given mode is calculated by integrating Eq.~\eqref{eq:electric_field_normalized_equation} with the approximated induced nonlinear polarization in Eq.~\eqref{Polarization_comp}.

\subsection{Estimation of perturbative response}
\label{sec:methods_perturbative_response}

We estimate perturbative nonlinear susceptibilities by extracting pathway-specific contributions from the computed response via TDSE. Operating in close analogy to \cite{mukamel1995principles} Eq. (4.85), we adapt it for multiphoton processes:

\begin{equation}
    \chi_{p}(\omega_0) = \frac{\int_{\omega_0\pm\Delta\omega}d\omega \vec{E}^{*,p}(\omega ) \cdot \tilde{\vec{P}}^{p}_{NL}( \omega )}{\epsilon_0\int_{\omega_0\pm\Delta\omega}d\omega \vec{E}^{p}(\omega ) \cdot\vec{E}^{*, p}(\omega )}\,.
\end{equation}

Here, $\omega_0$ is the central angular frequency of the generated signal, $\Delta\omega$ denotes the bandwidth of the corresponding harmonic, and $p \in {\text{THG},\text{FWM},\text{SFG}}$ labels the nonlinear pathway. The pathway-resolved nonlinear polarization, $\tilde{\vec{P}}^{p}_{NL}(\omega)$, is isolated using a lock-in filter by varying the relative phase between the two driving fields.

The field factor $\vec{E}^{p}(\omega)$ represents the pathway-specific product of the driving fields appearing in Eqs.~\ref{achiral_response} and \ref{chiral_response}. For finite-duration pulses, however, the nonlinear response receives contributions from all frequency combinations within the pulse bandwidth that satisfy the corresponding mixing condition. To account for this effect, the products are constructed in the time domain and subsequently Fourier transformed. For example, the third-harmonic pathway is represented by

\begin{equation}
\vec{E}^{THG}_{\mu}(\omega)
=
FFT\left[\vec{E}^{3}_\mu(t)\right].
\end{equation}

Analogous expressions are used for the remaining pathways. Since the extracted susceptibilities depend nontrivially on the intensities of both the $\omega$ and $2\omega$ driving fields, the above procedure is repeated over a grid of fundamental and second-harmonic intensities, yielding intensity-dependent susceptibility maps.

\subsection{Perturbative capillary nonlinear propagation}
\label{sec:methods_perturbative_capillary}

For our perturbative capillary model, we assume the driving fields are in the TM$_0$ mode at $\omega_0$ and the TE$_1$ mode at $2\omega_0$. The induced nonlinear polarization then becomes: 

\begin{align}
    &\tilde{\vec{P}}_{\mathrm{NL}}(\vec{r},3\omega_0) \epsilon_0^{-1}
    =
    \chi_{\mathrm{THG}}
    \vec{E}_{TM_{0}}(\vec{r},\omega_0)^3
    \nonumber \\
    &\quad
    +
    \chi_{\mathrm{FWM}}
    \vec{E}_{TE_{1}}(\vec{r},2\omega_0)^2
    \vec{E}_{TM_{0}}^{*}(\vec{r},\omega_0)
    \nonumber \\
    &\quad
    +
    \chi_{\mathrm{SFG}}
    \bigg(
        \vec{E}_{TM_{0}}(\vec{r},\omega_0)
        \times
        \vec{E}_{TE_{1}}(\vec{r},2\omega_0)
    \bigg)\,.
    \label{eq:PNL-pathways-cap}
\end{align}

Inserting this polarization into the propagation equation Eq.~\eqref{eq:electric_field_normalized_equation}, the amplitude of a given target mode $\mu$ at the frequency $3\omega_0$ follows as:

\begin{align}
    c_\mu(3\omega_0, z ) = i \bigg[ c_{\text{TM}_0}(\omega_0)^3 \kappa_{\mu}^\text{THG} \int_0^z e^{i \Delta k_{\mu}^\text{THG} z'} dz' +\nonumber \\  
    c_{\text{TM}_0}(\omega_0)^* c_{\text{TE}_1}(2\omega_0)^2 \kappa_{\mu}^\text{FWM} \int_0^z e^{i \Delta k_{\mu}^\text{FWM} z'} dz' + \nonumber \\
    c_{\text{TM}_0}(\omega_0) c_{\text{TE}_1}(2\omega_0) \kappa_{\mu}^\text{SFG} \int_0^z  e^{i \Delta k_{\mu}^\text{SFG} z'}dz'  \bigg]
\,.\end{align}
Here the phase mismatch $\Delta k$ of the individual pathways is given by:

\begin{eqnarray}
    &&\Delta k_{\mu}^\text{THG} = 3 \beta_{\text{TM}_0}(\omega_0) - \beta_\mu(3\omega_0)\,, \\
    &&\Delta k_{\mu}^\text{FWM} = 2 \beta_{\text{TE}_1}(2\omega_0) - \beta_{\text{TM}_0}(\omega_0 )- \beta_\mu(3\omega_0)\,, \\ 
    &&\Delta k_{\mu}^\text{SFG} = \beta_{\text{TM}_0}(\omega_0) + \beta_{\text{TE}_1}(2\omega_0 ) - \beta_\mu(3\omega_0)\,.
\end{eqnarray}
Furthermore the nonlinear coupling constants $\kappa$ are given by

\begin{align}
    \kappa_{\mu}^\text{p} = \frac{3 \omega_0}{c N_\mu(3\omega_0)}  \frac{ \chi_\text{p} h_\mu^\text{p} }{\int_A | \vec{F}_\mu(\vec{r}_\perp, 3\omega_0)|^2 d\vec{r}_\perp  }\,,
\end{align}

where $\chi_\text{p}$ is the nonlinear susceptibility associated with pathway $p \in \{\text{THG, FWM, SFG}\}$, and $h_\mu^\text{p}$ is the mode-averaged pathway-specific correlation function:

\begin{eqnarray}
    &&h_\mu^\text{THG} = \int_A \vec{F}_{\text{TM}_0}(\vec{r}_\perp, \omega_0)^3\cdot \vec{F}_\mu(\vec{r}_\perp, 3 \omega_0)^* d\vec{r}_\perp\,, \\
    &&h_\mu^\text{FWM} = \\
    &&\int_A \vec{F}_{\text{TE}_1}(\vec{r}_\perp, 2\omega_0)^2 \vec{F}_{\text{TM}_0}(\vec{r}_\perp, \omega_0)^* \cdot \vec{F}_\mu(\vec{r}_\perp, 3\omega_0)^* d\vec{r}_\perp\,, \nonumber \\ 
    &&h_\mu^\text{SFG} = \label{eq:specific_h3} \\&& \int_A \bigg( \vec{F}_{\text{TM}_0}(\vec{r}_\perp, \omega_0) \times \vec{F}_{\text{TE}_1}(\vec{r}_\perp, 2\omega_0)\bigg)  \cdot \vec{F}_\mu(\vec{r}_\perp, 3\omega_0)^* d\vec{r}_\perp \,.\nonumber  
\end{eqnarray}

Note that Eq.\eqref{eq:specific_h3} generalizes to Eq.\eqref{corr_function} in case of arbitrary driving modes.
Due to the modal dispersion in the waveguide the phase matching for all pathways is maximized in the TM$_2$ mode, and thus the intensity carried by the waveguide follows:

\begin{align}
    \text{I}(z,3\omega_0) \sim 
    |c_{\text{TM}_2}(z, 3\omega_0) |^2 \,.
\end{align}

Defining the parameters

\begin{eqnarray}
    &&\eta_\text{THG} = \frac{c_{\text{TM}_0}(\omega_0)^3 \kappa_{\text{TM}_2}^\text{THG}}{\Delta k_{\text{TM}_2}^\text{THG}} \,,\\
    &&\eta_\text{FWM} = \frac{ c_{\text{TM}_0}(\omega_0)^* c_{\text{TE}_1}(2\omega_0)^2 \kappa_{\text{TM}_2}^\text{FWM}}{\Delta k_{\text{TM}_2}^\text{FWM}}\,,\\
    &&\eta_\text{SFG} = \frac{ c_{\text{TM}_0}(\omega_0) c_{\text{TE}_1}(2\omega_0) \kappa_{\text{TM}_2}^\text{SFG}}{\Delta k_{\text{TM}_2}^\text{SFG}}\,.
\end{eqnarray}
The intensity can be written as:
\begin{eqnarray}
    \text{I}&&(z,3\omega_0) \sim  \bigg| \eta_\text{THG} ( e^{i \Delta k_{\text{TM}_2}^\text{THG} z }  -1 ) +\nonumber \\  
    &&\eta_\text{FWM} ( e^{i \Delta k_{\text{TM}_2}^\text{FWM} z }  -1 ) + 
    \eta_\text{SFG} ( e^{i \Delta k_{\text{TM}_2}^\text{SFG} z }  -1 ) \bigg |^2\,,
    \label{modal_intensity}
\end{eqnarray}

where we have assumed $\Delta k_\mu ^p \neq 0$.

\subsection{Lock-in detection}
\label{sec:methods_Lock-in}

Given the coexistence of of three pathways, a local field at frequency $3\omega$, even at the output or farfield, can be written as: 

\begin{align}
    \vec{E}_\text{loc}(\phi_{2\omega,\omega}) = \vec{E}_\text{THG} +\vec{E}_\text{SFG}e^{i\phi_{2\omega,\omega}} +\vec{E}_\text{FWM}e^{i2\phi_{2\omega,\omega}} \,.
\end{align}

where $\vec{E}_p$ are the field amplitudes generated by the respective pathway $p$ and $\phi_{2\omega,\omega}$ is the relative phase between the fundamental and the second harmonic. 
The local intensity then follows: 

\begin{align}
    I_\text{loc}(\phi_{2\omega,\omega})  = \tilde{I}_0 + \tilde{I}_1  e^{i\phi_{2\omega,\omega}}  + 
    \tilde{I}_1 e^{i2\phi_{2\omega,\omega}} + c.c. \,,
    \label{eq:lockin_inten}
\end{align}

where $c.c.$ stands for the complex conjugate and refers to adding the complex conjugate of the entire right hand side of the equation. The lock-in amplitudes $\tilde{I}_i$, where $i$ is the channel number are given by:

\begin{align}
    \tilde{I}_0 =& \frac{1}{2} \big(  |\vec{E}_\text{FWM}|^2 + |\vec{E}_\text{THG}|^2 + |\vec{E}_\text{SFG}|^2 \big)\,, \\
    \tilde{I}_1 =& \big( \vec{E}_\text{SFG} \cdot \vec{E}_\text{FWM}^* + \vec{E}_\text{THG} \cdot \vec{E}_\text{SFG}^* \big)\,,\\
    \tilde{I}_2=&\vec{E}_\text{THG} \cdot \vec{E}_\text{FWM}^* \,.
\end{align}

Lock in detection, i.e., recording data as a function of $\phi_{2\omega,\omega}$ and Fourier transformation with respect to it, is a direct measurement of the complex lock in amplitudes. This allows the isolation of the enantio-sensitive terms leading $e^{i\phi_{2\omega,\omega}}$ into lock-in channel 1 which is directly proportional to $\vec{E}_\text{SFG}$.
If the polarization of the nonlinear signals is parallel, and exempt the unlikely and easily avoidable case that the two terms in $\tilde{I}_1$ exactly, channel 1 encodes molecular handedness while channel 0 gives access to molecular concentration and channel 2 records the relative phase and amplitude of THG and FWM emission which in itself can serve as a achiral, molecule specific fingerprint.\\

Even in the case of parallel polarizations the lock-in measurement by itself does not suffice for a unique (up to common phase) reconstruction of the pathway specific nonlinear amplitudes and phases, because it only establishes relative amplitude difference between THG and FWM. For a fully unique reconstruction the amplitude of one of the achiral pathways has to be measured in isolation, e.g. measuring THG intensity by blocking the second harmonic driving field. Given independently measured $|E_\text{THG}|$ and complex lock-in amplitudes the complex pathway resolved amplitudes and their relative phases can be uniquely reconstructed through:

\begin{align}
    E_\text{FWM} =& \frac{\tilde{I}^*_2}{E_\text{THG}^*}\,,\\
    E_\text{SFG} =& \frac{E_\text{FWM}\tilde{I}_1  -  E_\text{THG}\tilde{I}^*_1}{|E_\text{FWM}|^2  - |E_\text{THG}|^2}\,,
\end{align}

where $E_\text{THG}$ carries the unknown common phase. 

\subsection{Perturbative nonlinear propagation in free space foci}
\label{sec:methods_free_space}

To avoid any errors from paraxial effects we choose to calculate the focal fields of the radially and azimuthally focused vector beams using the Richards--Wolf formalism \cite{richards1959electromagnetic,youngworth2000focusing}. The initially transverse polarized collimated vector beams enter an aplanatic focusing system satisfying the sine condition, so that the radial coordinate $\rho_i$ in the entrance pupil is mapped to the polar angle $\theta$ of the focused spherical wave according to
\begin{align}
    \rho_i = f \sin\theta ,
\end{align}
with $\theta \in [0,\theta_{\max}]$. The maximum focusing angle is set by the numerical aperture, $\mathrm{NA}=n\sin\theta_{\max}$.

The Richards--Wolf field in the focal region is
\begin{align}
    &\vec{E}(\rho,\phi,z)
    =\nonumber
    i\frac{k f}{2\pi} 
    \int_0^{2\pi}
    \int_0^{\theta_{\max}} \\&\quad
    \vec{U}(\phi',\theta)
    e^{i k z \cos\theta}
    e^{i k \rho \sin\theta \cos(\phi-\phi')}
    \sin\theta \, d\theta \, d\phi' .
    \label{eq:RW-specialized}
\end{align}
Here $\vec{U}(\phi',\theta)$ is the vectorial field on the reference sphere after the focusing element. The factor $\sqrt{\cos\theta}$ accounts for the apodization of the aplanatic system.

For a radially polarized input beam, the transverse pupil field is
\begin{align}
    \vec{A}_{\mathrm{rad}}(\rho_i,\phi')
    =
    a_{01}(\rho_i)
    \begin{pmatrix}
        \cos\phi' \\
        \sin\phi'
    \end{pmatrix},
    \label{eq:pupil-radial}
\end{align}
where $a_{01}$ is the radial amplitude profile of the radially polarized LG$_{01}$ mode, while the 2D polarization vector is radially for all $\phi'$. After the focusing element, the polarization vector is tilted into the local meridional plane, giving
\begin{align}
    \vec{U}_{\mathrm{rad}}(\phi',\theta)
    =
    a_{01}(f\sin\theta)\sqrt{\cos\theta}
    \begin{pmatrix}
        \cos\theta \cos\phi' \\
        \cos\theta \sin\phi' \\
        -\sin\theta
    \end{pmatrix}.
    \label{eq:U-radial}
\end{align}
Substitution into Eq.~\eqref{eq:RW-specialized} and analytical integration over $\phi'$ gives
\begin{align}
    \vec{E}_{\mathrm{rad}}(\rho,\phi,z)
    =
    -k f
    \begin{pmatrix}
        \cos\phi \, R_\perp^{01}(\rho,z) \\
        \sin\phi \, R_\perp^{01}(\rho,z) \\
        i R_z^{01}(\rho,z)
    \end{pmatrix},
    \label{eq:E-radial-LG01}
\end{align}
with
\begin{align}
    &R_\perp^{01}(\rho,z)
    =
    \int_0^{\theta_{\max}} \nonumber \\&\quad
    a_{01}(f\sin\theta)
    \sqrt{\cos\theta}
    \sin\theta \cos\theta
    J_1(k\rho\sin\theta)
    e^{i k z \cos\theta}
    d\theta ,
    \label{eq:Rperp01}
    \\
    &R_z^{01}(\rho,z)
    =
    \int_0^{\theta_{\max}}\nonumber \\&\quad
    a_{01}(f\sin\theta)
    \sqrt{\cos\theta}
    \sin^2\theta
    J_0(k\rho\sin\theta)
    e^{i k z \cos\theta}
    d\theta .
    \label{eq:Rz01}
\end{align}
Thus a radially polarized LG$_{01}$ beam develops a sizable longitudinal electric-field component upon tight focusing \cite{dorn2003sharper}. On the optical axis, $J_1(0)=0$ and $J_0(0)=1$, so the transverse field vanishes while the longitudinal field remains finite.\\

For an azimuthally polarized input beam, the transverse pupil field is
\begin{align}
    \vec{A}_{\mathrm{az}}(\rho_i,\phi')
    =
    a_{11}(\rho_i)
    \begin{pmatrix}
        -\sin\phi' \\
        \cos\phi'
    \end{pmatrix},
    \label{eq:pupil-azimuthal}
\end{align}
where $a_{11}$ is the radial amplitude profile of the azimuthally polarized LG$_{11}$ mode, while the polarization vector is azimuthal for all $\phi'$. Since the azimuthal polarization is perpendicular to the local meridional plane, the focusing transformation does not generate a longitudinal component. The field on the reference sphere is therefore
\begin{align}
    \vec{U}_{\mathrm{az}}(\phi',\theta)
    =
    a_{11}(f\sin\theta)\sqrt{\cos\theta}
    \begin{pmatrix}
        -\sin\phi' \\
        \cos\phi' \\
        0
    \end{pmatrix}.
    \label{eq:U-azimuthal}
\end{align}
Insertion into the Richards--Wolf integral yields
\begin{align}
    \vec{E}_{\mathrm{az}}(\rho,\phi,z)
    =
    k f A_\phi^{11}(\rho,z)
    \begin{pmatrix}
        \sin\phi \\
        -\cos\phi \\
        0
    \end{pmatrix},
    \label{eq:E-azimuthal-LG11}
\end{align}
where
\begin{align}
    &A_\phi^{11}(\rho,z)
    =\nonumber\\&
    \int_0^{\theta_{\max}}
    a_{11}(f\sin\theta)
    \sqrt{\cos\theta}
    \sin\theta
    J_1(k\rho\sin\theta)
    e^{i k z \cos\theta}
    d\theta .
    \label{eq:Aphi11}
\end{align}
The tightly focused azimuthally polarized LG$_{11}$ beam therefore remains purely transverse and azimuthally polarized.

The relevant input amplitudes of Laguerre Gaussian beams may be written, up to an overall normalization, as
\begin{align}
    a_{p\ell}(\rho_i)
    \propto
    \left(\frac{\sqrt{2}\rho_i}{w_i}\right)^{|\ell|}
    L_p^{|\ell|}\left(\frac{2\rho_i^2}{w_i^2}\right)
    \exp\left(-\frac{\rho_i^2}{w_i^2}\right),
    \label{eq:LG-amplitude-general}
\end{align}
where $w_i$ is the beam radius in the pupil plane. Thus the LG$_{01}$ and LG$_{11}$ cases are obtained by inserting the corresponding radial profiles $a_{01}(f\sin\theta)$ and $a_{11}(f\sin\theta)$ into Eqs.~\eqref{eq:Rperp01}--\eqref{eq:Aphi11}.\\

The local nonlinear induced polarization in the focus is obtained via the extracted pathway-resolved susceptibilities:
\begin{align}
    &\tilde{\vec{P}}_{\mathrm{NL}}(\vec{r},3\omega_0)
    =
    \chi_{\mathrm{THG}}
    \vec{E}_{LG_{01}}(\vec{r},\omega_0)^3
    \nonumber \\
    &\quad
    +
    \chi_{\mathrm{FWM}}
    \vec{E}_{LG_{11}}(\vec{r},2\omega_0)^2
    \vec{E}_{LG_{01}}^{*}(\vec{r},\omega_0)
    \nonumber \\
    &\quad
    +
    \chi_{\mathrm{SFG}}
    \bigg(
        \vec{E}_{LG_{01}}(\vec{r},\omega_0)
        \times
        \vec{E}_{LG_{11}}(\vec{r},2\omega_0)
    \bigg).
    \label{eq:PNL-pathways}
\end{align}
Here $\vec{E}_{LG_{01}}$ and $\vec{E}_{LG_{11}}$ are the focal fields obtained through numerical evaluation of Eq. \eqref{eq:RW-specialized}.

\subsubsection{Generated field from the nonlinear polarization}

The generated field at $3\omega_0$ is obtained from the vector Helmholtz equation. In a homogeneous medium with refractive index $n_3$ at the generated frequency,
\begin{align}
    \nabla \times \nabla \times
    \tilde{\vec{E}}_{\mathrm{gen}}(\vec{r},3\omega_0)
    -
    k_3^2
    \tilde{\vec{E}}_{\mathrm{gen}}(\vec{r},3\omega_0)
    =
    k_{0,3}^2
    \frac{
        \tilde{\vec{P}}_{\mathrm{NL}}(\vec{r},3\omega_0)
    }{\epsilon_0},
    \label{eq:vector-helmholtz-PNL}
\end{align}
where
\begin{align}
    k_{0,3}=\frac{3\omega_0}{c},
    \qquad
    k_3=n_3 k_{0,3}.
\end{align}
Using the outgoing dyadic Green function $\mathbf{G}(\vec{r}-\vec{r}\,')$, the solution can be written as
\begin{align}
    \tilde{\vec{E}}_{\mathrm{gen}}(\vec{r},3\omega_0)
    =
    \frac{k_{0,3}^2}{\epsilon_0}
    \int
    \mathbf{G}(\vec{r}-\vec{r}\,')
    \tilde{\vec{P}}_{\mathrm{NL}}(\vec{r}\,',3\omega_0)
    d^3r' .
    \label{eq:Egen-dyadic-Green}
\end{align}
In free space or a homogeneous isotropic medium, the dyadic Green function is
\begin{align}
    \mathbf{G}(\vec{r})
    =
    \left[
        \mathbf{I}_3
        +
        \frac{1}{k_3^2}
        \nabla\nabla
    \right]
    \frac{e^{ik_3 r}}{4\pi r},
    \label{eq:dyadic-green}
\end{align}
where $\mathbf{I}_3$ is the $3\times 3$ identity matrix. 

\subsubsection{Angular far field of the generated radiation}

In the radiation zone, the dyadic Green function reduces to the transverse projection of the Fourier-transformed nonlinear polarization. For an observation point
\begin{align}
    \vec{R}=R\hat{\vec{s}},
\end{align}
with
\begin{align}
    \hat{\vec{s}}
    =
    \begin{pmatrix}
        \sin\Theta \cos\Phi \\
        \sin\Theta \sin\Phi \\
        \cos\Theta
    \end{pmatrix},
\end{align}
the radiating wave vector is
\begin{align}
    \vec{k}_3(\Theta,\Phi)
    =
    k_3 \hat{\vec{s}} .
\end{align}
Only the transverse part of the nonlinear polarization contributes to the emitted far field:
\begin{align}
    \tilde{\vec{P}}_{\perp}(\Theta,\Phi)
    =
    \left[
        \mathbf{I}_3
        -
        \hat{\vec{s}}\hat{\vec{s}}^{\mathrm{T}}
    \right]
    \tilde{\vec{P}}_{\mathrm{NL}}
    \left(
        k_3\hat{\vec{s}},3\omega_0
    \right).
    \label{eq:PNL-transverse-projection}
\end{align}
The far field is therefore
\begin{align}
    \tilde{\vec{E}}_{\mathrm{gen}}(R,\Theta,\Phi)
    =
    \frac{k_{0,3}^2 e^{ik_3 R}}{4\pi\epsilon_0 R}
    \left[
        \mathbf{I}_3
        -
        \hat{\vec{s}}\hat{\vec{s}}^{\mathrm{T}}
    \right]
    \tilde{\vec{P}}_{\mathrm{NL}}
    \left(
        k_3\hat{\vec{s}},3\omega_0
    \right).
    \label{eq:Egen-farfield}
\end{align}
The angular intensity is proportional to
\begin{align}
    I_{\mathrm{gen}}(\Theta,\Phi)
    \propto
    \left|
        \tilde{\vec{E}}_{\mathrm{gen}}(R,\Theta,\Phi)
    \right|^2 .
\end{align}
The projection operator in Eq.~\eqref{eq:Egen-farfield} enforces the transversality condition
\begin{align}
    \hat{\vec{s}}\cdot
    \tilde{\vec{E}}_{\mathrm{gen}}(R,\Theta,\Phi)
    =
    0.
\end{align}

Equivalently, the far field may be parameterized by the transverse wave-vector components
\begin{align}
    k_x &= k_3\sin\Theta\cos\Phi,\\
    k_y &= k_3\sin\Theta\sin\Phi,\\
    k_z &= +\sqrt{k_3^2-k_x^2-k_y^2}.
\end{align}
The positive square root describes radiation propagating in the positive $z$ direction. The corresponding unit wave vector is
\begin{align}
    \hat{\vec{k}}
    =
    \frac{1}{k_3}
    \begin{pmatrix}
        k_x\\
        k_y\\
        k_z
    \end{pmatrix}.
\end{align}
In this notation,
\begin{align}
    \tilde{\vec{E}}_{\mathrm{gen}}(k_x,k_y)
    =
    \frac{k_{0,3}^2 e^{ik_3 R}}{4\pi\epsilon_0 R}
    \left[
        \mathbf{I}_3
        -
        \hat{\vec{k}}\hat{\vec{k}}^{\mathrm{T}}
    \right]
    \tilde{\vec{P}}_{\mathrm{NL}}
    \left(
        k_x,k_y,k_z,3\omega_0
    \right).
    \label{eq:Egen-kxky}
\end{align}

\subsubsection{Refocusing through the collimating lens}

The generated angular field is collected and refocused by the collimating lens. For an aplanatic collection system, the lens maps the emission angle $\Theta$ onto the radial coordinate $\rho_c$ in the collimated output plane according to
\begin{align}
    \rho_c=f_c\sin\Theta ,
    \label{eq:collection-sine}
\end{align}
where $f_c$ is the focal length of the collimating lens. The collection aperture restricts the field to
\begin{align}
    0\leq \Theta \leq \Theta_{\mathrm{col}},
    \qquad
    \mathrm{NA}_{\mathrm{col}}=n_3\sin\Theta_{\mathrm{col}} .
\end{align}
The field immediately after the collimating lens is
\begin{align}
    \vec{A}_{\mathrm{gen}}(\rho_c,\Phi)
    =
    B_{\mathrm{col}}(\rho_c)
    \frac{1}{\sqrt{\cos\Theta}}
    \hat{P}^{-1}_{x,y}(\Phi,\Theta)
    \tilde{\vec{E}}_{\mathrm{gen}}(\Theta,\Phi),
    \label{eq:collimated-generated-field}
\end{align}
with
\begin{align}
    \Theta
    =
    \arcsin\left(\frac{\rho_c}{f_c}\right).
\end{align}
The collection pupil is
\begin{align}
    B_{\mathrm{col}}(\rho_c)
    =
    1-
    H\left(
        \rho_c
        -
        f_c\sin\Theta_{\mathrm{col}}
    \right).
\end{align}
The factor $1/\sqrt{\cos\Theta}$ reverses the aplanatic apodization of the spherical wave, and $\hat{P}^{-1}_{x,y}$ maps the local transverse far-field polarization back onto the transverse Cartesian basis of the collimated beam.

\subsubsection{Near-field propagation by angular-spectrum propagation}

The generated near field can be calculated by propagating the angular spectrum produced by the nonlinear polarization through the focal volume. First, the nonlinear polarization is Fourier transformed in the transverse coordinates at each axial source position $z'$:
\begin{align}
    &\tilde{\vec{P}}_{\mathrm{NL}}(k_x,k_y,z',3\omega_0)
    =\nonumber\\&\quad
    \iint
    \tilde{\vec{P}}_{\mathrm{NL}}(x,y,z',3\omega_0)
    e^{-i(k_xx+k_yy)}
    dxdy .
    \label{eq:PNL-transverse-spectrum}
\end{align}
For each transverse wave vector,
\begin{align}
    k_z(k_x,k_y)
    =
    \sqrt{k_3^2-k_x^2-k_y^2}.
    \label{eq:kz-angular-spectrum}
\end{align}

Using the angular-spectrum representation of the outgoing Green function, the generated field at the observation plane $z$ is
\begin{align}
    &\tilde{\vec{E}}_{\mathrm{gen}}(k_x,k_y,z)
    =\nonumber\\&
    \frac{i k_{0,3}^2}{2\epsilon_0}
    \int
    \frac{1}{k_z}
    \left[
        \mathbf{I}_3
        -
        \hat{\vec{k}}_{\pm}\hat{\vec{k}}_{\pm}^{\mathrm{T}}
    \right]
    \tilde{\vec{P}}_{\mathrm{NL}}(k_x,k_y,z',3\omega_0)
    e^{i k_z |z-z'|}
    dz' .
    \label{eq:angular-spectrum-volume-propagation}
\end{align}
Here
\begin{align}
    \hat{\vec{k}}_{\pm}
    =
    \frac{1}{k_3}
    \begin{pmatrix}
        k_x\\
        k_y\\
        \pm k_z
    \end{pmatrix},
\end{align}
where the sign indicates forward and backward propagating wave respectively. Since we only need the forward propagating field, we obtain
\begin{align}
    &\tilde{\vec{E}}_{\mathrm{gen}}^{(+)}(k_x,k_y,z)
    = \nonumber\\&
    \frac{i k_{0,3}^2}{2\epsilon_0 k_z}
    \int_{-\infty}^{z}
    \left[
        \mathbf{I}_3
        -
        \hat{\vec{k}}_{+}\hat{\vec{k}}_{+}^{\mathrm{T}}
    \right]
    \tilde{\vec{P}}_{\mathrm{NL}}(k_x,k_y,z',3\omega_0)
    e^{i k_z (z-z')}
    dz' .
    \label{eq:angular-spectrum-forward}
\end{align}

The real-space near field is obtained by the inverse transverse Fourier transform,
\begin{align}
    \vec{E}_{\mathrm{gen}}(x,y,z)
    =
    \frac{1}{(2\pi)^2}
    \iint
    \tilde{\vec{E}}_{\mathrm{gen}}(k_x,k_y,z)
    e^{i(k_xx+k_yy)}
    dk_x dk_y .
    \label{eq:nearfield-inverse-transform}
\end{align}
This angular-spectrum propagation retains diffraction, interference between different axial source planes, and the vectorial polarization structure of the generated field. The far-field expression in Eq.~\eqref{eq:Egen-farfield} is recovered by evaluating the propagated field on the radiation shell and taking the asymptotic limit $R\rightarrow\infty$.

For the collected far field, only propagating components are retained:
\begin{align}
    k_x^2+k_y^2 \leq k_3^2 .
\end{align}
Components satisfying
\begin{align}
    k_x^2+k_y^2 > k_3^2
\end{align}
are evanescent. They may contribute to the immediate near field but do not propagate to the collection optics.

\subsection{Perturbative experimental parameters}
\label{sec:perturbative_parameters}
\label{sec:methods_perturbative_parameters}

The pulse, focusing, capillary, and medium parameters used in the perturbative estimates and in the confinement comparison of Sec.~\ref{sec:confinement_advantage} are summarized in Table~\ref{tab:parameters}.

\begin{center}
\centering
\small
\captionof{table}{Experimental and material parameters used in the calculation.}
\label{tab:parameters}
\setlength{\tabcolsep}{3pt}%
\resizebox{\columnwidth}{!}{%
\begin{tabular}{@{}lll@{}}
\toprule
\textbf{Parameter} & \textbf{Symbol} & \textbf{Value} \\
\midrule

\multicolumn{3}{@{}l}{\textbf{Pulse parameters}} \\
Fundamental wavelength & $\lambda_{\omega}$ & \SI{800}{\nano\meter} \\
Second-harmonic wavelength & $\lambda_{2\omega}$ & \SI{400}{\nano\meter} \\
Fundamental pulse energy & $E_{\omega}$ & \SI{1.06}{\micro\joule} \\
Second-harmonic pulse energy & $E_{2\omega}$ & \SI{0.59}{\micro\joule} \\
Intensity FWHM duration & $\tau$ & \SI{30}{\femto\second} \\

\midrule
\multicolumn{3}{@{}l}{\textbf{Focal parameters}} \\
Input numerical aperture & $\mathrm{NA}_{\mathrm{in}}$ & 0.085 \\
Input focal length & $f_{\mathrm{in}}$ & \SI{150}{\milli\meter} \\
Output numerical aperture & $\mathrm{NA}_{\mathrm{out}}$ & 0.085 \\
Output focal length & $f_{\mathrm{out}}$ & \SI{150}{\milli\meter} \\
Fundamental pupil diameter & $D_{4\sigma,\mathrm{pupil},\omega}$ & \SI{5.34}{\milli\meter} \\
Second-harmonic pupil diameter & $D_{4\sigma,\mathrm{pupil},2\omega}$ & \SI{5.34}{\milli\meter} \\
Fundamental focus diameter & $D_{4\sigma,\mathrm{focus},\omega}$ & \SI{57.2}{\micro\meter} \\
Second-harmonic focus diameter & $D_{4\sigma,\mathrm{focus},2\omega}$ & \SI{57.2}{\micro\meter} \\
Fundamental peak intensity & $I_{\mathrm{peak},\mathrm{focus},\omega}$ & \SI{2.01}{\tera\watt \centi\meter^{-2}} \\
Second harmonic peak intensity & $I_{\mathrm{peak},\mathrm{focus},2\omega}$ & \SI{2.11}{\tera\watt \centi\meter^{-2}} \\

\midrule
\multicolumn{3}{@{}l}{\textbf{Capillary parameters}} \\
Capillary diameter & $d_{\mathrm{cap}}$ & \SI{70}{\micro\meter} \\
Fundamental TM$_0$ diameter& $D_{4\sigma,\mathrm{cap},\omega}$ & \SI{57.2}{\micro\meter} \\
Second Harmonic TE$_1$ diameter& $D_{4\sigma,\mathrm{cap},2 \omega}$ & \SI{57.2}{\micro\meter} \\
Fundamental peak intensity & $I_{\mathrm{peak},\mathrm{cap},\omega}$ & \SI{2.0}{\tera\watt \centi\meter^{-2}} \\
Second harmonic peak intensity & $I_{\mathrm{peak},\mathrm{cap},2\omega}$ & \SI{2.0}{\tera\watt \centi\meter^{-2}} \\
\midrule
\multicolumn{3}{@{}l}{\textbf{Medium parameters}} \\
Gas pressure & $p$ & \SI{50}{\milli\barunit} \\
Temperature & $T$ & \SI{404}{\kelvin} \\
Linear susceptibility at $\omega$ & $\chi^{(1)}_{\omega}$ & $2.92 \times 10^{-5}$ \\
Linear susceptibility at $2\omega$ & $\chi^{(1)}_{2\omega}$ & $3.52 \times 10^{-5}$ \\
Linear susceptibility at $3\omega$ & $\chi^{(1)}_{3\omega}$ & $4.37 \times 10^{-5}$ \\
Chiral SFG susceptibility
& $\chi^{(2)}_{\mathrm{ch}}$
& $(1.07 - 1.04i)\times 10^{-17}\,\mathrm{m\,V^{-1}}$ \\
Achiral THG susceptibility
& $\chi^{(3)}_{\omega+\omega+\omega}$
& $(2.02 + 0.53i)\times 10^{-26}\,\mathrm{m^2\,V^{-2}}$ \\
Achiral FWM susceptibility
& $\chi^{(3)}_{2\omega+2\omega-\omega}$
& $(-3.11 + 0.87i)\times 10^{-26}\,\mathrm{m^2\,V^{-2}}$ \\

\bottomrule
\end{tabular}%
}
\end{center}

\bibliographystyle{apsrev4-2}
\bibliography{lit} 

\section {Acknowledgments}
Funded by the European Union (ERC, ULISSES, 101054696). Views and opinions expressed are however those of the author(s) only and do not necessarily reflect those of the European Union or the European Research Council. Neither the European Union nor the granting authority can be held responsible for them. P.V.Z. is grateful to the Simons Foundation for the computational resources. D.A. aknowledges Funding from Royal Society URF R251036. A.R. and L.R. acknowledge funding from “la Caixa” Foundation (ID 100010434), under the agreement “LCF/BQ/PR24/12050018”, from the Spanish Ministry of Science, Innovation and Universities \& the State Research Agency through the project ref. PID2024-163024NA-I00 (MICIU/AEI/10.13039/501100011033/FEDER, UE), and support from the Severo Ochoa Centers of Excellence program through Grant CEX2024-001445-S.

\section {Author Contributions}
A.L. conceived the study. A.L. and J.T. performed simulations. P. V.-Z. and D.A. contributed the DFT dipole inputs for simulations. S. P. contributed quantum chemistry expertise. A.R. and L.R. contributed to discussions on free space propagation. M.I. and O.S. supervised the study. A.L. and O.S. wrote the first draft of the paper.
\end{document}